\documentclass[]{constructions}

\newcommand{\correspondingauthor}{Céline Mougenot, Senior Lecturer, Associate Professor, Imperial College London}

\begin{document}

\thispagestyle{plain}

\newcolumntype{Y}{>{\raggedleft\arraybackslash}X}

 \begin{flushright}
 \small \textit{March 2026} \\
 \end{flushright}


\vspace{30pt}
\begin{center}
    \LARGE{\textbf{YAQIN: Culturally Sensitive, Agentic AI for Mental Healthcare Support Among Muslim Women in the UK}}
\end{center}

\begin{center}
\vspace{4pt}
\large
    Yasmin Zaraket, 
    Dr. Céline Mougenot\textsuperscript{1}
    
\small
   Dyson School of Design Engineering, Imperial College London 

\end{center}

\begin{small}
\begin{center}
\vspace{9pt}
\textbf{Abstract}    
\end{center}

\begin{adjustwidth}{20pt}{20pt}
\label{abstract}

Mental healthcare services in the UK lack the tools and the resources to meet nuanced cultural needs of Muslim women to avoid them feeling their values are pathologised and, consequently, miss opportunities of trust and engagement~\cite{bignall2020racial}. 
Despite growing awareness of cultural competency, few interventions meaningfully integrate Islamic frameworks into therapeutic support. 
This report investigates the design and evaluation of YAQIN, a co-designed AI-based application that supports cultural and faith-sensitive mental health engagement for Muslim women. 
With the population of Muslim women in England increasing annually from almost $1.9$ million in 2021, YAQIN responds to a significant gap in care~\cite{ONS2023ReligionAgeSex}.
It leverages AI's anonymity, continuous support, and contextual intelligence to offer a faith-aware chatbot and guided journaling tool, both grounded in user-centred design principles where Islamic interpretations of psychology are key.

The YAQIN Design Research Methodology encompasses three iterative stages. 
The first was rigorous contextual investigation and review of available literature.
Second, initial user research included semi-structured interviews with N=$14$ stakeholders, comprising both Muslim women who had sought mental health services and mental health experts. 
YAQIN analyses the insights through deductive thematic analysis and translated them into personas, journey maps, and design specifications which informed its first prototypes.

This project conducted the evaluation through a co-designed user study involving five participants: four Muslim women who engaged in a live interaction and co-design session, and one mental health expert who provided targeted critique on therapeutic alignment and cultural sensitivity after using the chatbot prototype. 
YAQIN assessed the design of the chatbot based on tone, faith relevance, and emotional resonance based on key co-design feedback focused on personalised reflection prompts, Quranic contextualisation, and the Retrieval-Augmented Generation pipeline used for contextual continuity.

Participants praised its ability to bridge cultural gaps, particularly in trust, articulation, and therapeutic confidence.
YAQIN successfully fosters emotional reflection, validates faith-based identity, and supports therapy readiness as evidenced by the evaluation findings.
Moreover, feedback highlighted areas for growth, such as supporting linguistic diversity and offering deeper routine-based guidance.
YAQIN's design and development implemented this feedback into the final iteration.

This project leverages the value of culturally-sensitive AI to improve mental healthcare accessibility and trust for marginalised communities. 
It contributes to current systems by operationalising Islamic psychological conceptual in digital care tools, highlighting the broader potential of contextual faith-integrated, user-informed technology in healthcare innovation.


\end{adjustwidth}

\end{small}

\vspace{10pt}
 \section{Introduction}
 \label{sec:intro}
 \label{cp:introduction}

Mental healthcare resources in the UK are strained and barely meet basic demand, leaving little room for nuanced services and cultural accommodations. 
Challenges such as systemic bias, cultural stigma, and the lack of religiously informed therapeutic models contribute to potential disengagement from mainstream services~\cite{bignall2020racial, nhsrho2022rapidreview}. 
Despite increased recognition of cultural competency in healthcare, many services lack the training, time, and resources needed to meaningfully accommodate faith-sensitive perspectives~\cite{haque2016integrating}. 

This project addresses the disconnect and focuses on bridging the gap between conventional therapy and {\em the needs of Muslim women} by exploring how recent advances in artificial intelligence (AI) and innovation can support mental healthcare interventions with cultural sensitivity. 
In $2021–22$, only $2.6\%$ of almost $3.8$ million Muslims in England, with over $1.9$ million women included, completed their referred mental healthcare plans~\cite{bma2025mentalhealth, ONS2023ReligionAgeSex}, highlighting a critical need for more inclusive and faith-aligned pathways to mental healthcare.

This project develops and utilises {\em novel} culturally and faith-sensitive {\em agentic AI} to bridge the disconnect between Muslim women and mainstream mental healthcare, specifically focusing on client user experience (UX).
While initiatives like Muslim Youth Helpline~\cite{MYH2025}, Inspirited Minds~\cite{InspiritedMinds2025}, and the Muslim Women's Network UK~\cite{MWNUK2025}, offer valuable community support, they are often limited to set hours and constrained by volunteer capacity~\cite{MYH2025, InspiritedMinds2025}.
YAQIN's use of AI complements these efforts by offering scalable, on-demand, and personalised support.

Termed YAQIN, 
\textarabic{يقين} 
Arabic for certainty, it incorporates a Retrieval-Augmented Generation (RAG) pipeline 
that connects Islamic knowledge, Islamic interpretations of psychology, and novel journaling features within a chatbot agentic interface. 
This ensures contextual continuity and dynamically surfaces relevant and culturally assuring knowledge during client interactions. 

{\bf Motivation:~}Consider Lina, a student struggling with exam anxiety, unable to explain to her therapist how supplication helps her feel calmer and less anxious. 
YAQIN offered a bridge by reframing supplications as breathing-based recitation exercises with meaningful reassuring phrases. This improved communication and engagement with her therapist.

\begin{figure}[b!]
    \centering
    \includegraphics[width=\linewidth]{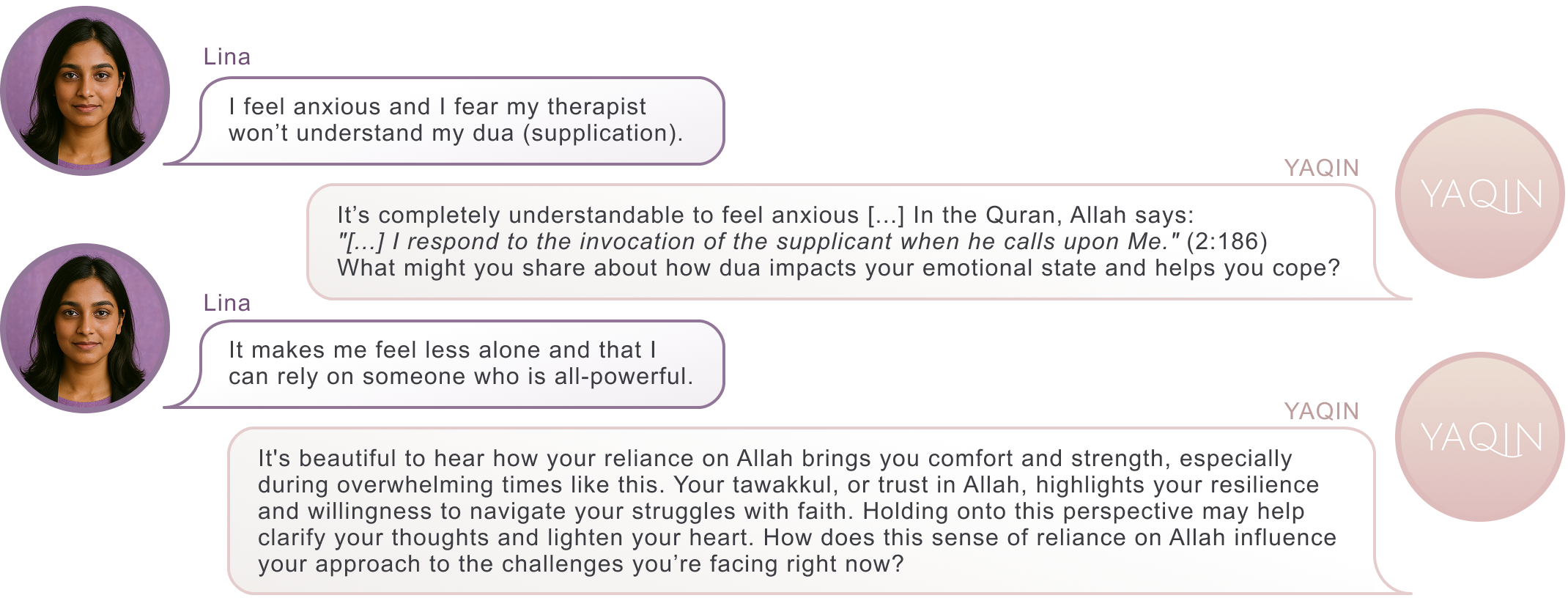}
    \caption{Motivating Use Case with Lina Persona}
    \label{fig:motivation}
\end{figure}

The central research question is:

\textit{How might we co-design a culturally and faith-sensitive AI tool, informed by Islamic resources, to support Muslim women navigating mental healthcare in the UK?}

YAQIN targets cultural minority groups and focuses on UK-based Muslim women as a first success story. 
It is an innovative, culturally rooted, AI-powered application built to validate and support users' cultural, emotional and spiritual needs. 
Its RAG presents a contextualisation framework to include cultural sensitivity and leverages Islamic interpretations of psychology to bridge an observable gap in mental healthcare for Muslim women.
This framework is adaptable to other similar contexts and service gaps.

The YAQIN Design Research Methodology (DRM) framework
combines literature review, semi-structured interviews with seven Muslim women and seven mental healthcare experts, and deductive thematic analysis in the Descriptive Study I stage, 
concept and initial prototype development in the Prescriptive Study stage, 
and evaluation, iterative prototyping, and discussion reflecting on broader implications and future potential in the Descriptive Study II stage.
They appear in Sections ~\ref{cp:literatureReview}-~\ref{cp:discussion}, respectively.

Insights from each stage translate into user personas, journey maps, and design specifications, informing the development of YAQIN.
In the Descriptive Study II stage, four participants joined a co-design session and one expert participated in usability testing to assess the YAQIN chatbot therapeutic relevance, emotional resonance, and usability.

In summary, YAQIN, implemented, successfully deployed, and tested on Huggingface, with a click-through UI, responds to a pressing need for inclusive, empathetic, and contextually grounded mental health solutions.
By leveraging faith-sensitive psychological paradigms and user-centred design, YAQIN revolutionises culturally-sensitive AI tools in therapeutic engagement. 
It contributes to a broader conversation on how healthcare technology can be ethically and critically aligned with the lived realities of marginalised communities.

 \section{Project Aim \& Objectives}
 \label{sec:aims}
 \label{cp:Background&Objectives}

This project aims to explore how AI can support culturally and faith-sensitive mental health care. 
By focusing on Muslim women's experiences, it develops an ethical, spiritually grounded agentic intervention that enhances therapeutic trust and emotional validation through contextually relevant, Islamically informed design and interaction.

\noindent
\textbf{Objectives:}
\begin{itemize}
    \item Understand emotional and cultural barriers Muslim women face accessing mental healthcare.
    \item Identify themes, needs, and aspirations on the relevance, trust, and emotional experiences of Muslim women through in-depth stakeholder interviews.
    \item Develop user personas and journey maps grounded in interview qualitative data synthesis via deductive thematic analysis.
    \item Map the pain points and highlighted insights to opportunities in leveraging emerging technology like Large Language Models (LLM) and agentic AI integrations.
    \item Design and prototype YAQIN, integrating agentic AI to address identified needs.
    \item Evaluate YAQIN effectiveness in fostering cultural sensitivity and competence in therapeutic contexts.
\end{itemize}

\section{Literature Review}
\label{sec:review}

\label{cp:literatureReview}

Cultural competency is an emerging concept in mental healthcare.
Statutory services often feel ``monocultural'' and lacking religious competence, particularly for ethnic minority communities who do not see their values or worldviews reflected in mainstream provision~\cite{bansal2022understanding}.
These concerns are echoed by both the NHS Race and Health Observatory (RHO) and the British Medical Association (BMA), who note that mental health services remain under-resourced in addressing the needs of ethnically diverse communities~\cite{nhsrho2022rapidreview, bma2025mentalhealth}.
As a result, many individuals from ethnic minority backgrounds avoid care due to mistrust and fear of racist or culturally uninformed treatment~\cite{nhsrho2022rapidreview, bma2025mentalhealth}.

The Muslim Mind Collaborative (MMC) has argued that the ``need for culturally and faith-sensitive'' mental healthcare must shift from a specialist concern to a mainstream clinical priority~\cite{mmc2024calltoaction}.
Recent data illustrate the impact of this gap: in $2021-2022$, only 2.6\% of Muslim clients completed therapy plans, compared to 18.4\% of Christians and 38.9\% of non-religious clients~\cite{bma2025mentalhealth}.
Focusing on UK Muslims, $44\%$ of them felt faith-related issues were inadequately addressed in mainstream counselling~\cite{bma2025mentalhealth}.
A key factor is lack of religious inclusivity. 
Ahmad et al. argue that integrating Islamic perspectives can bridge important gaps in therapy~\cite{ahmad2023muslimpatients} and improve mental healthcare outcomes~\cite{citystgeorges2022muslimmentalhealth}.

Reviews of cultural competence in mental healthcare warn that static, checklist-based models often fail to capture the fluid, intersectional realities of service users and may unintentionally realise cultural stereotypes~\cite{bhui2007cultural}.
These concerns support arguments that cultural responsivity in technology-enabled mental health systems must move beyond generic notions of ``cultural competence'' toward ongoing, context-aware adaptation that recognises users’ identities, histories, and power relations~\cite{eustis2023culturalTech}.
Narrative reviews of AI for positive mental health further highlight that, while AI systems can improve access, monitoring, and personalisation, these benefits will not be equitably distributed unless cultural and contextual factors are explicitly embedded into system design and evaluation~\cite{thakkar2024positiveAI}.

\subsection{Cultural Inaccessibility and Systemic Barriers for Ethnic Minority Women}

The Race Equality Foundation (REF) reports that women from diverse ethnic backgrounds are ``disproportionately affected by adverse mental health outcomes and underutilise services'' due to a combination of systemic discrimination, cultural inaccessibility, and mistrust~\cite{bignall2020racial}.
Muslim women in particular may feel alienated in therapy settings when their cultural norms and values are pathologised, minimised, or dismissed~\cite{bansal2022understanding}.
Fears around judgement, stigma, shame, and breaches of confidentiality further inhibit engagement with mental healthcare services~\cite{mmc2024calltoaction, citystgeorges2022muslimmentalhealth}.

Recent work in AI mental health design highlights how these barriers can be amplified or alleviated trrough technology.
The SMILE platform, for example, illustrates how an AI-driven mental health system can embed multilingual support and culturally sensitive therapeutic tools to reach diverse populations, explicitly positioning cultural adaptation as the centre of design rather than an optional layer~\cite{smile2025platform}.
Cross-cultural validation studies of AI-based mental health assessments show that linguistic nuances, culturally specific symptom expressions, and biased training data can lead to systematic misdiagnosis or under-recognition of distress among minoritised users~\cite{crossCulturalAI2025}.
Building on these concerns, overviews of AI in mental health warn that unless gender, race, migration status, and faith-based experiences are deliberately represented in datasets and outcome measures, AI systems risk reinforcing the existing invisibility of women and ethnic minorities in both access to and the quality of mental healthcare~\cite{thakkar2024positiveAI}.

\noindent\textbf{Identified themes:} \textit{Confidentiality and Trust; Changes in Access and Acceptance}

\subsection{Breaking Trust: Bias, Misinterpretation, and Cultural Disconnect}

Trust is particularly difficult to establish where Muslim women’s experiences are shaped by culturally uninformed care and provider bias~\cite{almutairi2022iraqi}.
Studies with Iraqi Muslim women in clinical settings report frequent experiences of being misunderstood, especially when norms around modesty and communication are overlooked or challenged by clinicians~\cite{almutairi2022iraqi}.
Healthcare staff sometimes question Muslim women's credibility, reinforcing feelings of marginalisation in clinical spaces~\cite{sherif2021impact}.
Muslim perspectives on mental illness typically integrate spiritual, social, and clinical dimensions, whereas dominant clinical frameworks often adopt secular biomedical models that foreground pathology over lived meaning~\cite{elahi2023ethnic}.
Muslim mental health practitioners therefore advocate embedding cultural and religious understanding within NHS care practices~\cite{tannerah2024consultations}.

Among Muslim minorities, stigma and a lack of cultural competence within services contribute to the underutilisation of mental healthcare~\cite{hussain2022mental}.
This can worsen academic and personal wellbeing, particularly for university students who may experience heightened cultural dissonance and fear of judgement~\cite{mahmud2024addressing}.
Many Muslim women fear that disclosing personal struggles could harm family reputation and social standing, discouraging them from seeking help or being open in counselling sessions~\cite{chen2023understanding}.
In this context, building therapeutic trust requires cultural competence, empathetic engagement, and respectful professional relationships that transform clinical settings from spaces of scrutiny into ``safe circles of healing.''

These issues are mirrored in debates about AI in mental health.
Ugar and Malele argue that there should not be a homogeneous or generic design for AI and machine learning systems diagnosing mental health disorders, because such systems embed value-laden judgements about what constitutes disorder that differ across cultures~\cite{ugar2024designingAI}.
They show that mental health diagnosis depends on both ``factual evidence'' and culturally situated value judgements, so AI systems trained primarily on Western notions of pathology may misclassify culturally normative experiences and undermine trust in regions such as sub-Saharan Africa~\cite{ugar2024designingAI}.
Broader work on cultural responsivity in technology-enabled services similarly warns that generic cultural competence frameworks can obscure within-group differences, and that trust depends on acknowledging users' specific histories, identities, and structural marginalisation~\cite{eustis2023culturalTech}.
Extending this insight, early research on “cultural prompting’’ for AI mental health agents shows that explicitly instructing large language models to attend to users’ cultural backgrounds can enhance perceived empathy and appropriateness of responses, offering a tangible mechanism for rebuilding trust in AI-mediated support~\cite{culturalPrompting2025}.

\noindent\textbf{Identified themes:} \textit{Therapist and Client Prejudice; Stigma and Cultural Taboo; Creating a Safe Environment}

\subsection{Faith-Aware Therapeutic Adaptations and Inclusive Clinical Practice}

Improving cultural competence is associated with stronger therapeutic relationships and better outcomes for Muslim clients~\cite{ali2023understanding}.
Rahman et al. emphasise the importance of ``truly culturally sensitive care,'' advocating for faith-informed, client-centred approaches that treat religious identity as a core part of the person rather than an optional add-on~\cite{rahman2022mental}.
Within many Muslim communities, stigma around ``family shame’’ complicates disclosure in therapeutic contexts, reinforcing concerns about confidentiality and moral judgement~\cite{chen2023understanding}.
Suggested adaptations include gender-concordant therapists and space for spiritual discussion when appropriate; Muslim clients often describe such arrangements as ``reassuring’’ and ``trust-building’’~\cite{saherwala2021providing}.

Spreitzer’s psychological empowerment framework highlights meaning, competence, autonomy, and impact as key dimensions across cultural contexts~\cite{spreitzer1995psychological}.
This framework reinforces the need to validate clients’ identities and provide choices that reflect their values, and it is particularly relevant for designing inclusive and responsive mental healthcare that mitigates religious and ethnic disparities.
Therapists also need to recognise that distress may be expressed through cultural or religious idioms-such as religious obsessive-compulsive disorder (OCD) or moral struggle~\cite{haque2016integrating}.
Language sensitivity, religious literacy, and familiarity with Islamic worldviews become critical for avoiding ruptures in the therapeutic relationship.

There are parallel developments in AI-mediated training and simulation that show how cultural competence can be supported through technology.
Co-design work on a chatbot for culturally competent clinical communication demonstrates that AI can simulate diverse patients and provide structured feedback guided by cultural competence models, helping trainees reflect on empathy, listening, and cross-cultural understanding, while also revealing current limitations such as the lack of non-verbal cues and overly agreeable virtual patients~\cite{radovanovic2025codesignChatbot}.
Behavioural health projects using AI and machine learning as deliberate-practice tools aim to improve clinicians’ cultural competence by giving targeted feedback on language, micro-aggressions, and responsiveness; here AI is positioned as a trainer helping them recognise and correct bias rather than a neutral replacement for therapists~\cite{sondermind2024culturalAI}.
Policy and ethics commentary on culturally appropriate care in the age of AI stresses that culture is flexible and evolving, and, therefore, AI systems and training tools must be designed to adapt to changing identities and community norms rather than freezing static stereotypes into algorithms~\cite{rand2024culturallyAppropriateAI}.

\noindent\textbf{Identified themes:} \textit{Improvement in Inclusivity and Representation; Language and Communication Barrier; Client Preferences; Conventional Models vs.\ Cultural Realities; Training, Supervision, and Peer Networks}

\subsubsection{Integrating Islamic Epistemologies Into Modern Mental Healthcare}

To adequately integrate religious and culturally sensitive care into therapy and counselling, it is essential to engage with Islamic interpretations of psychology and mental wellbeing.
Islamic psychology emphasises the \textit{nafs} (psyche) as eternal and spiritually anchored, contrasting with secular models that focus primarily on the mind as a cognitive or biochemical entity~\cite{haque2016integrating}.
Muslim clients often interpret mental distress through a sense of divine purpose, shaping their expectations about the role and limits of therapy~\cite{rahman2022mental}.
Culturally adapted CBT models illustrate how faith-based interpretations can be meaningfully aligned with evidence-based psychological practice~\cite{CollaborativeModel}.

The concept of \textit{tazkiyat al-nafs} (self-purification) encourages aligning the inner self with divine guidance to attain emotional and spiritual balance~\cite{haque2016integrating}.
Historical figures such as Abu Zayd al-Balkhi, a 9th-century scholar, distinguished between physical and psychological illness and proposed proto-cognitive therapeutic approaches, with significant convergence between his diagnostic criteria and contemporary frameworks like the DSM~\cite{haque2016integrating}.
He emphasised how thoughts influence mood and advocated preventative emotional care, making his work one of the earliest documented efforts to integrate psychotherapeutic insight within religious and medical frameworks~\cite{haque2016integrating}.

Contemporary Islamic mental healthcare frameworks build on these concepts.
Quranic teachings and religious coping practices-such as prayer (\textit{salat}), remembrance (\textit{dhikr}), and supplication (\textit{dua})-have been shown to reduce stress, anxiety, and depressive symptoms when integrated appropriately~\cite{raza2019mindfulness, owens2023interventions}.
Faith-sensitive assessment tools aligned with Islamic value systems support more accurate and culturally meaningful evaluation of mental states~\cite{haque2016integrating}.
However, there remains a shortage of culturally competent practitioners and limited time and resources in mainstream therapy~\cite{nhsrho2022rapidreview}.
These constraints highlight the need for scalable interventions, such as AI-assisted, faith-informed tools, that can augment both clients and practitioners in delivering culturally sensitive mental healthcare~\cite{ahmad2023impact}.

When AI systems interact with religious content, additional risks arise.
AI may misinterpret scripture or generate unauthoritative or misleading spiritual guidance, raising serious concerns among religious communities~\cite{AIreligion2024}.
There is particular worry that AI tools could dilute or bypass traditional sources of religious authority if not carefully constrained~\cite{AIreligion2024}.
Ethical discussions emphasise a strong requirement for AI to protect user privacy when handling spiritually sensitive disclosures, ritual practices, or matters of belief~\cite{AIreligion2024}.
When appropriately designed, AI can expand access to religious knowledge without substituting human scholars, but this requires explicit ethical guidelines to ensure authenticity, accuracy, and faith-aligned boundaries~\cite{AIreligion2024}.

\noindent\textbf{Identified themes:} \textit{Redefining Psychological Concepts with Islamic Terminology; Prophetic Stories and Quranic Anchors}

\subsubsection{Faith-Sensitive Design in Mental Healthcare}

Faith-sensitive approaches can help bridge critical gaps between Muslim clients and conventional mental healthcare models.
Islamic frameworks centred on the \textit{nafs} (self), \textit{qalb} (heart), and \textit{‘aql} (intellect) offer spiritually coherent models for understanding and treating emotional distress~\cite{haque2016integrating}.
When metaphysical constructs such as \textit{tawakkul} (trust in God) and \textit{sabr} (patience) are explicitly integrated into care, Muslim women often report increased trust in the therapeutic process and greater acceptance of interventions~\cite{haque2016integrating}.

Quran-based interventions such as structured recitation, guided reflection, and spiritual mentoring can support mental wellbeing, particularly when coupled with reflective journaling practices that help individuals clarify emotions and internal conflicts~\cite{smyth2018online, owens2023interventions}.
Therapy becomes more effective when Islamic narratives and literature are embedded within clinical interactions, shifting the model to serve the client’s worldview rather than forcing clients to bracket their faith.
In turn, this builds therapist-client relationships that respect both clinical needs and faith-based identities.

\noindent\textbf{Identified themes:} \textit{Integration of Religious Practices in Sessions; Faith-Sensitive Client-Centred Practice}

\subsection{The Role of AI in Expanding Culturally Sensitive Mental Healthcare}

To address gaps in cultural sensitivity and accessibility, this project examines how AI can support the mental healthcare needs of underserved cultural groups-particularly Muslim women-by embedding cultural and religious awareness into digital intervention design.

\subsubsection{Opportunities for Faith- and Culture-Aware AI Interventions}

Recent advances in multilingual conversational AI have opened access to nuanced cross-cultural knowledge that was previously difficult to operationalise at scale.
The UK Parliamentary Office of Science and Technology notes that ``AI-based tools could support healthcare professionals to provide personalised care more quickly,'' particularly for people underserved by traditional services~\cite{post2025ai}.
Ahmad et al.\ argue that a well-designed AI tool in mental healthcare can ``reduce systemic delay, ensure anonymity, and increase therapeutic access'' for underserved Muslim women in the UK~\cite{ahmad2023impact}.

Several interventions showcase how culturally sensitive design can be realised in practice.
The Chatbot-Assisted Self-Assessment (CASA) intervention, co-designed with ethnic minority communities, enables users to self-assess mental wellbeing and access tailored medical advice~\cite{nadarzynski2025chatbot}.
Tunggala proposes an Islamic ethics-based framework for aligning AI communication with religious values, offering a way to balance innovation with cultural and theological sensitivity~\cite{tunggala2025cross}.
In the broader app ecosystem, Calm Health, Headspace, and Wysa provide accessible and well-designed AI or wellness tools, but largely lack faith- and culture-sensitive framing~\cite{calmhealth, headspace, wysa}.
Faith-informed platforms such as Muslim Moodfit and Ruhi integrate Islamic content, yet do not offer robust personalisation or co-designed features~\cite{muslimmoodfit, ruhiapp}.
Inspirited Minds offers culturally aware therapy but does not currently include AI-based self-help or reflective tools~\cite{inspiritedminds}.
These gaps collectively motivate YAQIN’s focus on combining AI with co-designed, culturally and faith-sensitive support for Muslim women.

HCI work on tracking during Ramadan provides further insight into design needs.
Studies of Muslim women’s menstrual and religious tracking behaviours show frequent value conflicts between religious obligations (such as fasting) and physical or hormonal health needs, leading many users to “hack together’’ personal tracking workflows to align mainstream apps with fiqh-based rules, highlighting significant unmet design needs, as existing tools fail to represent nuanced religious and menstrual factors such as fasting exemptions or prayer interruptions, often resulting in stress, guilt, and self-judgment~\cite{ramadanTracking2024}.
At the interactional level, conversational agents have been shown to increase women’s participation and articulation of ideas in mixed-gender online discussions, suggesting that chatbots can amplify under-represented voices and mitigate social barriers tied to gender~\cite{PMC10477209}.

\noindent\textbf{Identified themes:} \textit{Brainstorming Features}

\subsubsection{Ethical, Cultural, and Technical Challenges}

Despite its promise, AI in mental healthcare introduces significant ethical and cultural challenges.
Emke warns that cultural insensitivity in translation algorithms may reinforce, rather than remove, linguistic and contextual errors, ``leading to a deterioration in output quality''~\cite{emke2024risk}.
AI models trained primarily on Western datasets may not generalise to culturally diverse users~\cite{dehbozorgi2025application}.
Data biases, privacy risks, and misinterpretation of user intent can render AI interventions exclusionary and harmful for vulnerable groups~\cite{johnson2024ai}.
YAQIN attempts to mitigate these issues by using co-design methods and restricting its responses to a curated semantic database, enabling retrieval-augmented generation (RAG) that is more culturally and theologically aligned~\cite{westminster2025chatbot}.

Research on culturally aware AI design shows that systems often fail when they do not meaningfully embed the cultural norms, communication styles, and value systems of their intended users~\cite{nelson2025culturalAI}.
This gap leads to biased data use, misinterpretation of cultural idioms, and outputs that can marginalise minority communities.
To avoid this, culturally aware AI must be developed through participatory design with the target community, ensuring the system reflects their lived experiences~\cite{nelson2025culturalAI}.

Effective localisation requires more than small interface changes; it involves adapting tone, language, examples, and reasoning patterns to the relevant cultural and religious context.
Ongoing community evaluation is also essential to maintain value alignment and prevent cultural erasure, ensuring cultural context shapes the model’s behaviour rather than only its surface features~\cite{nelson2025culturalAI}.

\noindent\textbf{Identified themes:} \textit{Communication and Language; Data and Privacy}

\subsection{Summary and Research Question}

Across this literature, there is clear evidence of persistent cultural, systemic, and religious misalignments between Muslim women and mainstream mental healthcare.
Barriers include institutional bias, stigma, fear of judgement, language challenges, and the absence of faith-integrated care.
At the same time, both clinical and technological literatures highlight the risks of culturally naive AI systems and the importance of participatory, context-aware, and faith-sensitive approaches.

With limited resources and structural inequalities, bridging these gaps requires both culturally informed therapeutic models and scalable innovations that address access, representation, and inclusivity at every stage of mental healthcare delivery.
This motivates the central research question of this project:

\begin{quote}
\textbf{\textit{How might we co-design a faith- and culturally-sensitive AI 
tool, informed by Islamic resources and personalised feedback to improve the relevance, trust, and emotional experiences of Muslim women navigating mental healthcare in the UK?}}
\end{quote}

 \section{Methodological Approach} 
 \label{sec:method}
 \label{cp:MethodologicalApproach}

\begin{figure}[tb!]
    \centering
    \includegraphics[width=\linewidth]{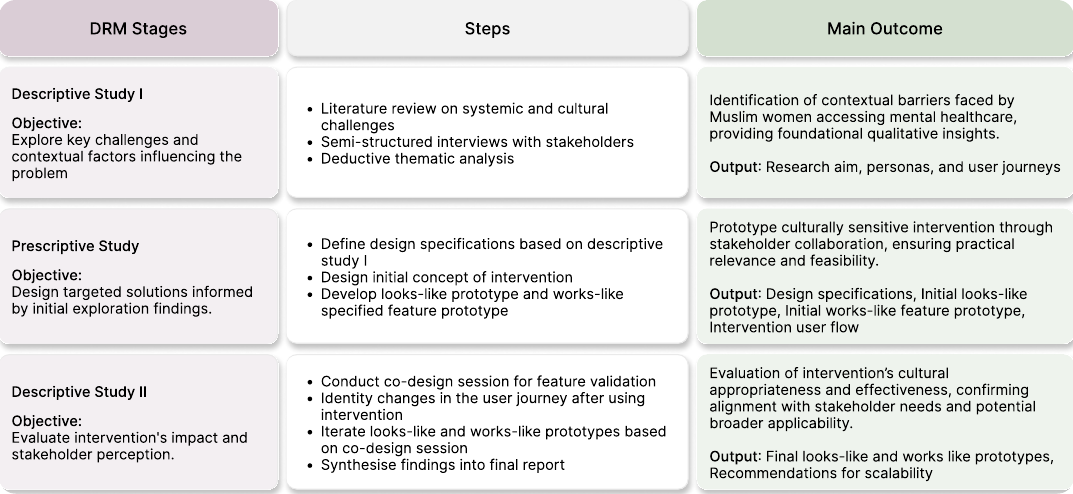}
    \caption{DRM Outline for YAQIN}
    \label{fig:DRM}
\end{figure}

This paper's methodological approach leverages the DRM~\cite{DRM}, which provides a structured yet flexible framework tailored for iterative research and the development of evidence-based design solutions.
DRM, illustrated in Figure~\ref{fig:DRM}, is characterised by three interconnected stages: Descriptive Study I, Prescriptive Study, and Descriptive Study II. 
Each stage strategically builds upon insights from the preceding stage and allows for refinement and robust validation of outcomes.

 \section{User Research \& Insights} 
 \label{sec:research}
 \label{cp:UserResearch&Insights}

This section presents primary user research following a literature review on cultural and religious factors in mental healthcare.

\subsection{Research Methodology}
This user research explores Muslim women's experiences navigating UK mental healthcare systems, aiming to enhance the relevance, trust, and emotional support. 

User interviews offer a powerful means of capturing nuanced, context-rich insights that cannot be accessed through literature alone~\cite{kvale2015interviews}. 
The interviews aim to:

\begin{enumerate}
    \item Deepen and validate insights from the literature by exploring the lived experiences of stakeholders.
    \item Identify key barriers Muslim women encounter when seeking mental healthcare and how trust can be built to overcome them.
    \item Understand the role of Islamic teachings and concepts in mental health experiences, and how these can be effectively integrated into therapy.
    \item Gather specific recommendations for designing culturally sensitive mental health interventions tailored for Muslim women.
\end{enumerate}

\subsubsection{Interview Structure \& Ethics}
 
This project conducted each $30–45$ minute interview in person or online over the course of three months, enabling a systematic exploration of culturally sensitive mental healthcare experiences.

The interviews focused on two primary stakeholder groups with the following inclusion criteria: 
\begin{enumerate}
    \item Muslim women in the UK (Clients) with experience accessing or consideration of mental healthcare services
    \item Mental healthcare psychotherapists or researchers (Experts)
\end{enumerate}

This project recruited interviewees via community networks and university events.
This may introduce sampling bias but it is important for this project to have this niche group of interviewees.
The average age for experts was $39.4$ years (SD = $12.0$), and for clients, $21.7$ years (SD = $1.1$).

Data saturation in homogenous groups is typically achieved with around $7$ interviews~\cite{Guest2020}.
This project conducted N=$14$ interviews with seven Muslim women and seven experts, ensuring thematic depth without oversampling.
To ensure wellbeing protection, interviewees were excluded if mental health discussions posed a risk of distress or emotional harm.
Table~\ref{tab:interviewees} lists the participating interviewees and their roles.

\begin{table}[tb!]
    \centering
    \caption{List of Interviewees and Their Occupation}
    \setlength{\extrarowheight}{4pt}
    \begin{tabular}{|p{55pt}|p{40pt}|p{40pt}|p{40pt}|p{40pt}|p{40pt}|p{40pt}|p{40pt}|}
        \hline
        \multirow{2}{55pt}{Muslim Women Clients} & 
        C1 & C2 & C3 & C4 & C5 & C6 & C7 \\
         & \scriptsize Under-graduate & \scriptsize Teacher & \scriptsize Medical Student & \scriptsize Under-graduate & \scriptsize Master's Student & \scriptsize Under-graduate & \scriptsize Under-graduate \\
         \hline
        \multirow{2}{55pt}{\\Mental Healthcare Experts} & 
        E1 & E2 & E3 & E4 & E5 & E6 & E7 \\
        & \scriptsize Psychologist, Psychotherapist & \scriptsize Pscho-therapist, Counsellor & \scriptsize Medical Researcher & \scriptsize Counsellor & \scriptsize Clinical Team Lead, Pscyho-therapist & \scriptsize Psychology Researcher, Islamic Scholar & \scriptsize Mental Health Pharmacist \\ \hline
    \end{tabular}
    \label{tab:interviewees}
\end{table}

The Imperial College Research Ethics Committee approved the ethics application prior to interviews. 
This project did not actively ask for personal mental health disclosure to minimise emotional risk or triggering.
Such disclosures would have required separate approval from the Science, Engineering and Technology Research Ethics Committee (SETREC).
The interviews began with verbal informed consent to record the session, knowing that they could withdraw at any time. 

\subsubsection{Deductive Thematic Analysis Approach}

Following Braun and Clarke's six-phase model of familiarisation, coding, theme development, review, definition, and reporting, the deductive thematic analysis method supports a systematic yet flexible exploration of meaning in qualitative data~\cite{Byrne2021}.

This project identifies key themes from Section~\ref{cp:literatureReview} like stigma, cultural barriers, trust in therapy, and integration of Islamic teachings that directly relate to the coding phase in deductive thematic analysis. 
Figure~\ref{fig:themes} below shows the codes grouped into broader thematic clusters. 
This structure grounded the interview analysis and validated literature insights through real-life narratives.

\begin{table}[tb!]
    \centering
    \caption{Thematic Analysis Code and Theme Breakdown from Literature Review}
    \begin{tabular}{|p{0.35\textwidth}|p{0.35\textwidth}|}
        \hline
        \textbf{Themes} & \textbf{Codes Identified in Literature Review} \\ \hline

        \multirow{2}{150pt}{Changes in Cultural and Religious Diversity Acceptance}
        & Improvement in Inclusivity and Representation \\ \cline{2-2}
        & Changes in Access and Acceptance \\ \hline

        \multirow{2}{150pt}{Stigma, Judgement, and Barriers to Care}
        & Stigma and Cultural Taboo \\ \cline{2-2}
        & Therapist and Client Prejudice \\ \hline

        \multirow{3}{150pt}{Islamic Knowledge in Therapy}
        & Integration of Religious Teachings in Sessions \\ \cline{2-2}
        & Prophetic Stories and Quranic Anchors \\ \cline{2-2}
        & Faith-Inclusive Client-Centred Practice \\ \hline

        \multirow{2}{150pt}{Current Clinical Practices and Framework Clashes}
        & Conventional Models vs Cultural Realities \\ \cline{2-2}
        & Training, Supervision, and Peer Networks \\ \hline

        \multirow{2}{150pt}{Trust and Safety in Therapy Sessions}
        & Confidentiality and Trust \\ \cline{2-2}
        & Creating a Safe Environment \\ \hline

        \multirow{3}{150pt}{Communication Barrier in Therapeutic Relationship}
        & Language and Communication Barrier \\ \cline{2-2}
        & Redefining Psychological Concepts with Islamic Terminology \\ \cline{2-2}
        & Client Preferences \\ \hline

        \multirow{3}{150pt}{Design Considerations for Mental Health Tools}
        & Communication and Language \\ \cline{2-2}
        & Brainstorming Features \\ \cline{2-2}
        & Data and Privacy \\ \hline
    \end{tabular}
    \label{fig:themes}
\end{table}

\subsection{Findings}

This project reveals consistent challenges and opportunities in culturally sensitive mental healthcare for Muslim women in the UK through deductive thematic analysis of user interviews. 
By engaging clients and experts, this analysis captures how cultural and religious values shape therapeutic engagement, expectations, and overall outcomes.

The tension between culturally rooted stigma and the need for non-judgmental and accessible  mental health support was the most consistent pain point. 
As one expert noted, 
\begin{quote}
    ``Historically, ethnic minorities do not reach out for mental health because it's not part of their kind of thinking [...] support is usually done within the community or spiritual community" (E5)
\end{quote} 
This validates the literature describing mental healthcare stigma as particularly pronounced in collectivist and religious communities where struggles may be dismissed as signs of weak faith or poor character.

Interviewees also highlighted the barriers to finding appropriate therapy. 
They expressed frustration with systems that felt unintuitive or culturally misaligned with barriers like cost, availability, and discomfort navigating conventional clinical norms. 
One client spoke of her struggle seeking help through the most affordable, accessible pathway; her university counselling service.
\begin{quote}
    ``The system wasn't really easy to access [...] there wasn't a clear place on the website to go to" (C3)
\end{quote}
 
While C3 ended up seeing a non-Muslim counsellor, she and the majority of interviewed clients expressed a preference for seeing one that's Muslim.
As C4 explained:  
\begin{quote}
    ``You can much more easily share [your struggles] with someone else who obviously grew up knowing about Islam or your culture at the same time as you" (C4)
\end{quote}
Both experts and clients noted that therapists of any background could offer effective care so long as they approached clients' faith as meaningful and valid. 
This supports literature advocating for ``cultural humility" over ``cultural matching".
Effective therapy relies on client-centricity.

Cultural mismatch was also evident when clients felt pressured to adopt therapeutic recommendations that conflicted with religious or cultural norms. 
As one client recalled:
\begin{quote}
``Let's say you're talking about how you're going to put boundaries with [...] your mum. The [therapist] would suggest: 'leave the house', 'go move out'. That's not an option that we want to consider, not only because [...] it's not Islamically the right thing to do. It doesn't make sense." (C2)
\end{quote}
This underscores the need for therapists to frame behavioural strategies within the bounds of clients' ethical and familial frameworks rather than imposing normative  values of individualism or autonomy.

Clients appreciated therapists who connected Islamic practices with psychological tools. 
Both clients and experts emphasised the power of integrating concepts like \textit{dua} (supplication), \textit{tawakkul} (leaving matters in God's hands), and  Prophet Muhammad's emotional experiences. 
One expert remarked,
\begin{quote}
"Even the Prophet suffered with sadness [...] there's nothing wrong with that" (E7),
\end{quote}
reframing emotional struggles as spiritually significant rather than shameful. 

Another expert invoked a Quranic verse to highlight the therapeutic potential of patience, saying,
\begin{quote}
``Take the verse ‘Faṣbir ṣabran jamīlan' [So endure patiently with a content patience (Quran 70:5)]. It acknowledges hardship yet commands [Muslims] to respond with \textit{sabr} [patience]. Recognising \textit{sabr} as a therapeutic skill reframes suffering as an opportunity for growth.'' (E6)
\end{quote}

One expert shared how integrating religious values supported a client with long-term depression. 
\begin{quote}
``[My] client described qualities she wished to cultivate like  \textit{ḥilm} (forbearance and abstinence), and I introduced a \textit{hadith} [(a narration)] from Imam Jafar al-Sadiq on the seven aspects of \textit{akhlaq} (personality ethics), linking spiritual ideals to therapeutic goals.'' (E1)
\end{quote}

\subsubsection{Persona \& User Journey Mapping}

The analysis uses personas and user journey maps, core user-centred design tools, to translate interviews into design specifications. 
Personas translate lived experiences into relatable profiles that support empathy-driven design~\cite{blomquist2002personas}, while journey maps reveal emotional, cultural, and systemic barriers over time~\cite{yale2023journey}. 
These methods align stakeholder needs with culturally-sensitive service design~\cite{springerHCD}.

\begin{figure}[tb!]
    \centering
    \begin{tabular}{c}
        \includegraphics[width=\linewidth]{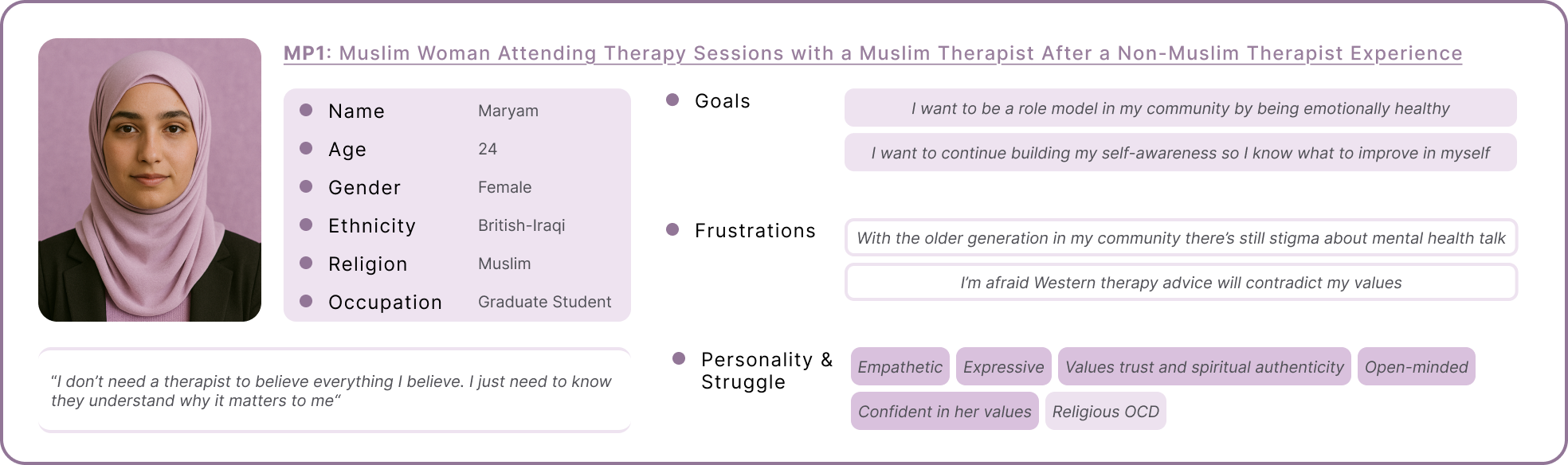} \\
        \includegraphics[width=\linewidth]{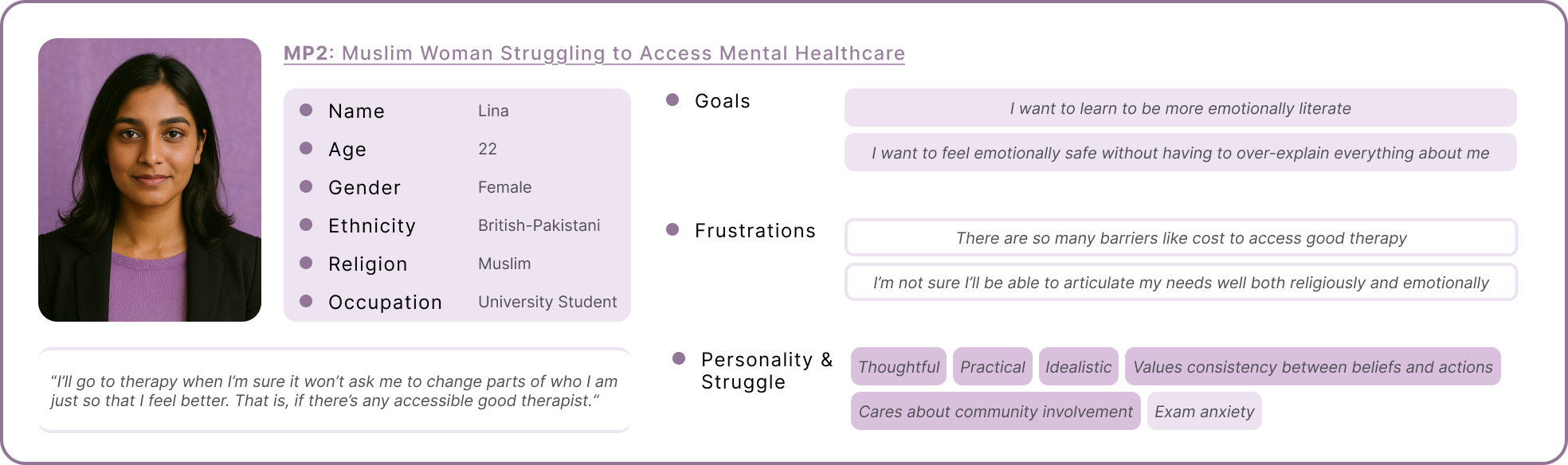} 
    \end{tabular}
    \caption{Personas (top) MP1 Maryam: Muslim Woman Attending Therapy Sessions with a Muslim Therapist After a Non-Muslim Therapist Experience and (bottom) MP2 Lina: Muslim Woman Struggling to Access Mental Healthcare~\cite{chatgpt}}
    \label{fig:MPpersonas}
\end{figure}

\begin{figure}[tb!]
    \centering
    \begin{tabular}{c}
        \includegraphics[width=\linewidth]{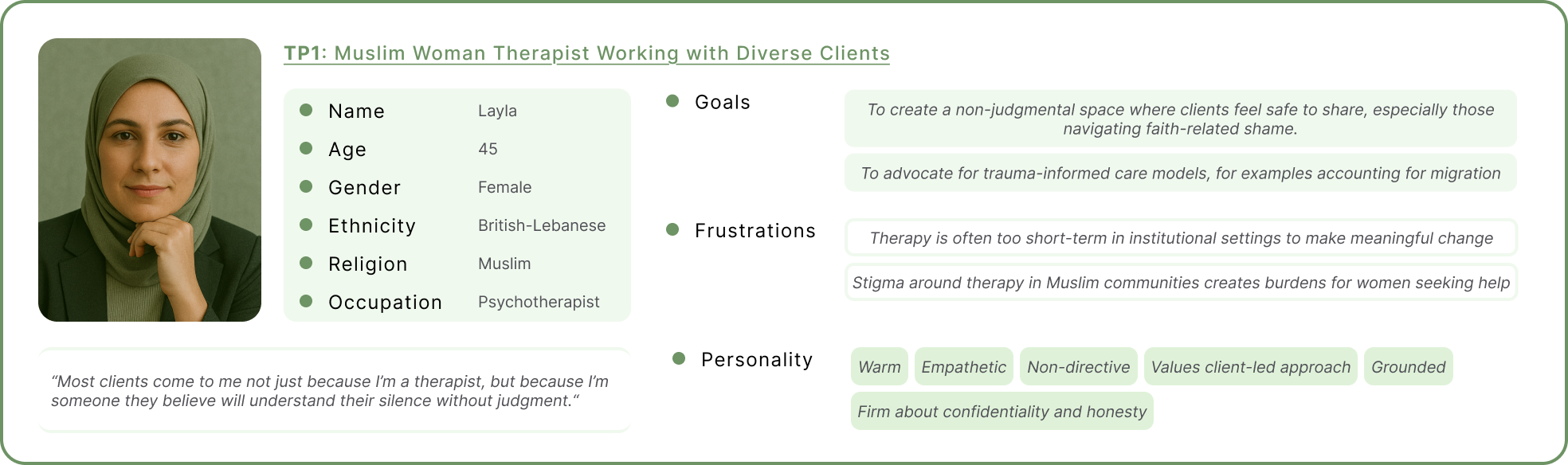} \\
        \includegraphics[width=\linewidth]{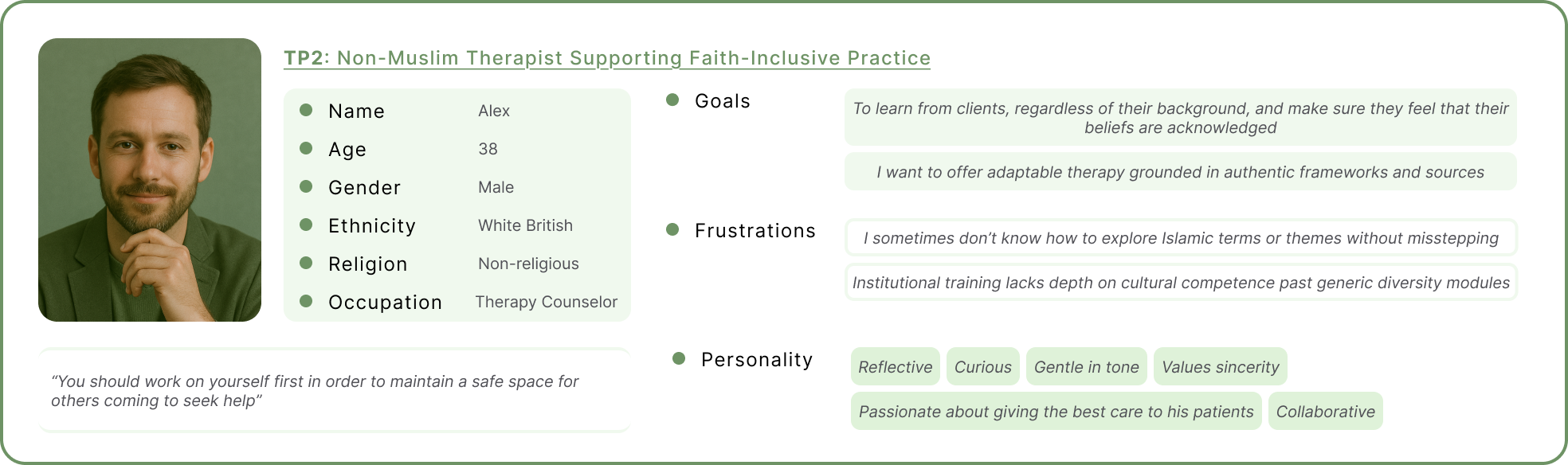}
    \end{tabular}
    \caption{Personas (top) TP1 Layla: Muslim Woman Therapist Working with Diverse Clients and (bottom) TP2 Alex: Non-Muslim Therapist Supporting Faith-Inclusive Practice~\cite{chatgpt}}
    \label{fig:TPpersonas}
\end{figure}

\begin{figure}[tb!]
    \includegraphics[width=\textwidth]{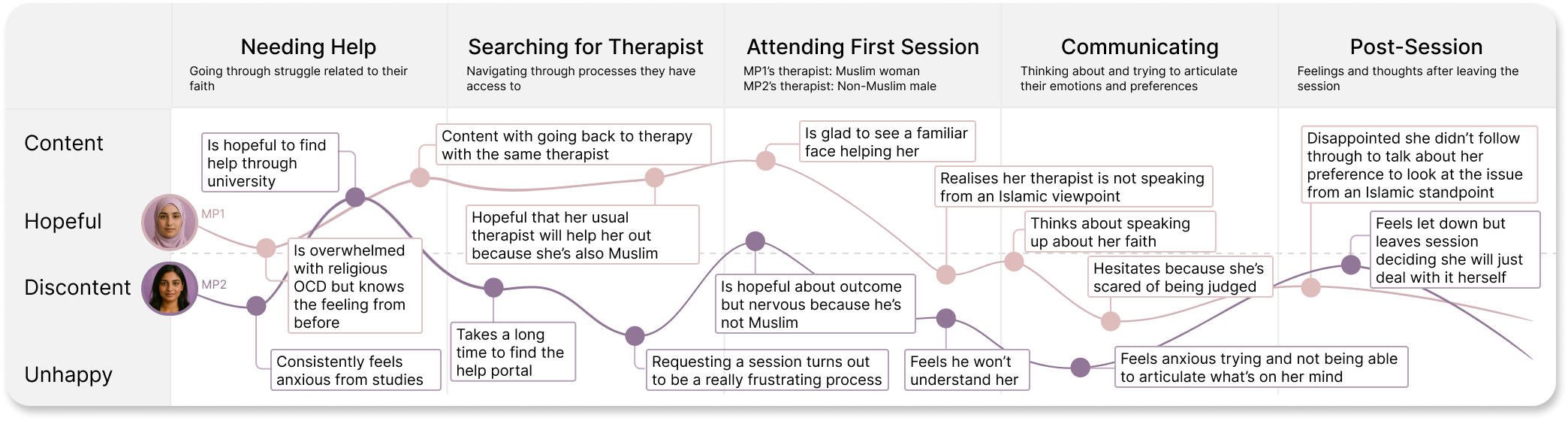}
    \caption{MP1 Maryam and MP2 Lina User Journey Maps: Different Journeys of Navigating Starting or Restarting Therapy Sessions}
    \label{fig:mpInitialjourney}
\end{figure}

\begin{figure}[tb!]
    \centering
    \includegraphics[width=\textwidth]{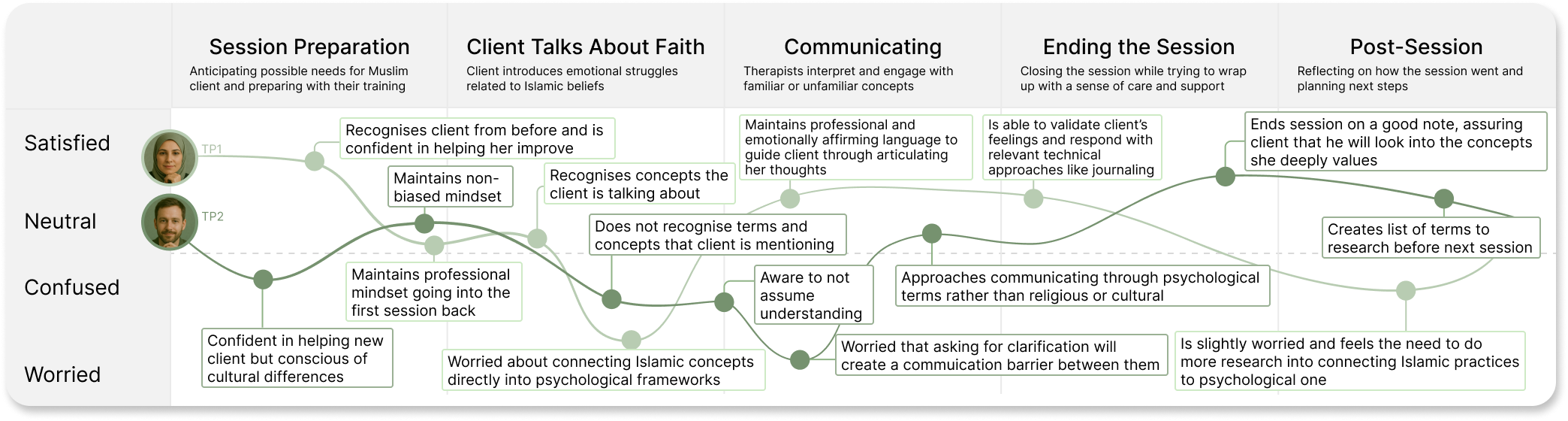}
    \caption{TP1 Layla and TP2 Alex User Journey Maps: Different Journeys of Navigating Clients with Familiar or Unfamiliar Cultural Backgrounds}
    \label{fig:tpInitialjourney}
\end{figure}

Insights from thematic analysis and contextual research informed the four personas in Figures~\ref{fig:MPpersonas} and~\ref{fig:TPpersonas} and their simplified journey maps in Figures~\ref{fig:mpInitialjourney} and~\ref{fig:tpInitialjourney}.

\subsection{Discussion}

The user journeys identify core pain points at every stage such as difficulty accessing culturally resonant care, fear of judgment, limited integration of faith, and emotional disconnects.
MP2's reluctance to continue therapy after feeling misunderstood or TP2's hesitation in engaging with unfamiliar religious language underscore a broader misalignment between current mental healthcare models and the realities of Muslim women seeking mental healthcare.

These pain points echo literature findings that the Muslim woman's experience in therapy is at risk of being culturally dissonant, inaccessible, or emotionally unsafe~\cite{blomquist2002personas, yale2023journey}. 
For instance, there is a large increase in clients' preference for therapists who inherently understand their values and struggles.
This accentuates the design need for safe ways to express faith (like \textit{dua} or Quranic reflection) without risking being pathologised. 

Issues such as emotional withdrawal during early therapy experiences, as seen in MP1's journey, point to a need for clear, culturally and religiously attuned communication.
Likewise, the desire for continued engagement outside therapy sessions backs the potential for structured, therapist-linked support tools that enhance therapy and do not replace it.

\subsubsection{Summary}

In summary, the user journeys and personas form the foundation for designing an intervention that reduces cultural friction, integrates meaningful Islamic constructs, and supports both emotional and therapeutic continuity. 
Each step in the journey should help shape YAQIN to align with the values, vulnerabilities, and healing aspirations of its users.

\subsubsection{Limitations}
Although the sample size of $14$ participants satisfies the saturation criteria, this project recognises that the findings may lack generalisation to Muslim populations outside the UK. 
While interviewees matched the defined target user-group, limited geographic and ethnic diversity narrowed the breadth of perspectives. 
This is a valuable consideration for future developments and broader application.
However, the focused scope enabled deep engagement with the target users, and their varied experiences offer strong directions for designing culturally sensitive interventions within the context of this project.

 \section{YAQIN App Design \& Development}
 \label{sec:design}
 \label{cp:des&dev}

This section details the features of YAQIN and the interacting components that realise them. 
They fulfill design specifications and embody target user values, concerns, and aspirations. They translate cultural sensitivity into user-centred app experiences.

\subsection{Design Specifications \& Success Criteria}

\begin{figure}[H]
  \centering
  \includegraphics[width=\textwidth]{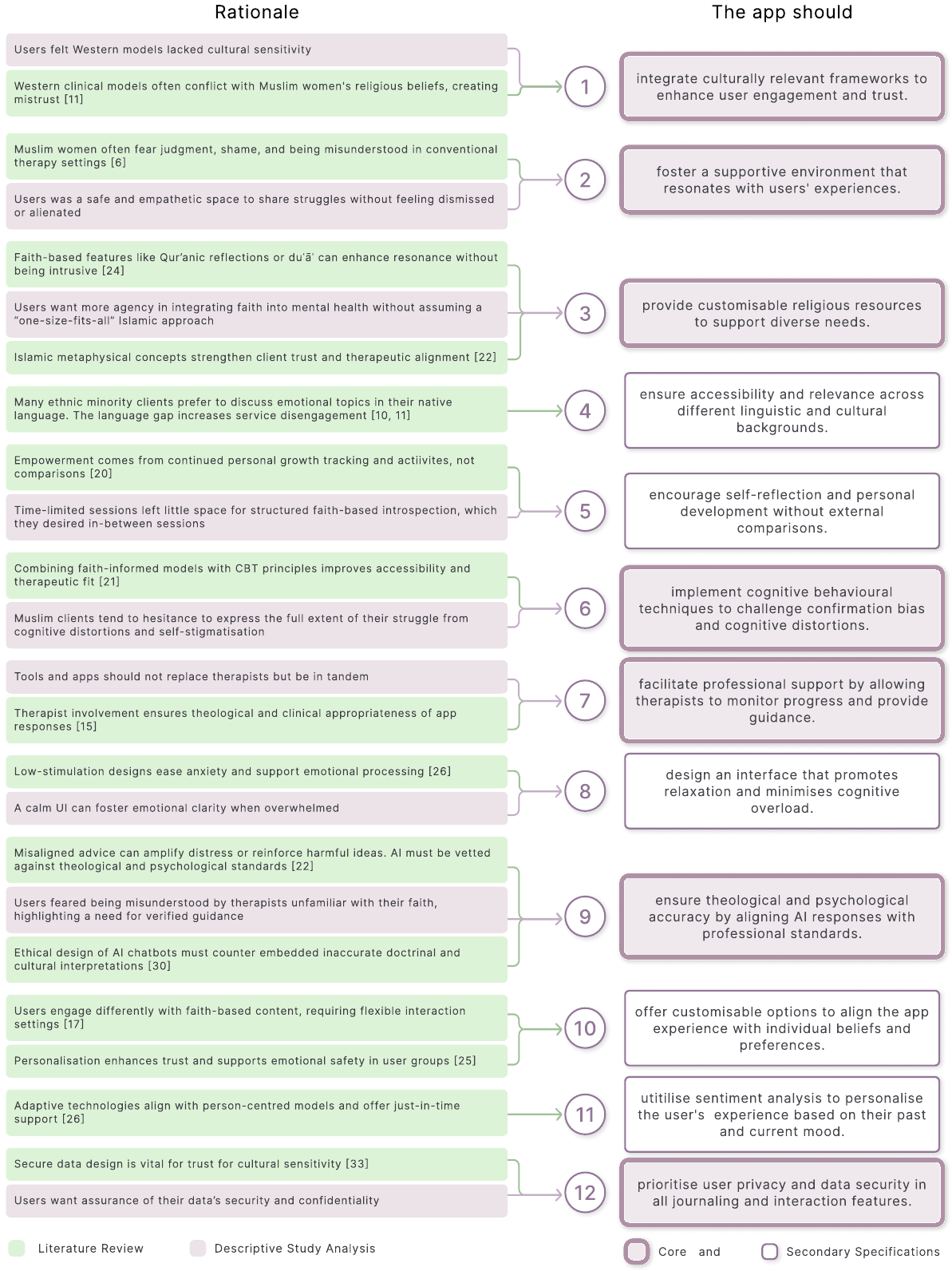}
  \caption{Design Specifications for YAQIN Intervention}
  \label{fig:desspec}
\end{figure}

YAQIN must satisfy the 
design specifications and their associated constraints to function effectively.
The descriptive study of Section~\ref{cp:UserResearch&Insights} results in these characteristics and constraints through the synthesis of user interviews, cultural analysis, and psychological literature. 
The resulting $12$ specifications show in Figure~\ref{fig:desspec} with their rationales. 
They include integrating culturally relevant paradigms and Islamic teachings, supporting cognitive behavioural strategies, and preserving data privacy.

\subsection{Ideation \& Concept Generation}

Participants frequently described misfits between Muslim moral and relational values and assumptions embedded in conventional mental healthcare practice. 
For example, advice to cut ties with family by therapists, is often religiously unacceptable, and culturally incompetent. 
One user summarised this discord: 
\begin{quote}
  ``It doesn't make sense [...] it's not Islamically the right thing to do” (C2).
\end{quote}

These misfits inspired \textit{YAQIN}, 
an AI-agentic application that bridges 
emotional and spiritual-cultural needs 
in a resource-strained mental healthcare system.
In this prescriptive study, 
the resulting looks-like and works-like prototypes focus on 
simulating
emotionally difficult conversations in a faith-sensitive, safe digital space. 

\subsubsection{Branding \& Visual Identity}

\textit{YAQIN} (\textarabic{يقين}) is an Arabic term that connotes a deep, unwavering sense of spiritual certainty. 
In epistemology, \textit{yaqin} reflects a tranquil and resolute state of cognition achieved through understanding oneself. 
The name reflects YAQIN's goals to support users through emotional confusion and align with their values. 
It implies clarity and healing can coexist with struggle and introspection.

The visual language of YAQIN as illustrated in Figure~\ref{fig:initialUIScreens} complements this epistemic grounding. 
A muted palette fosters psychological calm and reduces cognitive load on the interface~\cite{designandpaper2021muted}. 

\textbf{Design specifications met:} 1, 2, and 8.

\subsection{Main Feature Specification}

\begin{figure}[tb!]
  \centering
  \begin{tabular}{cccc}
    \includegraphics[width=0.22\linewidth]{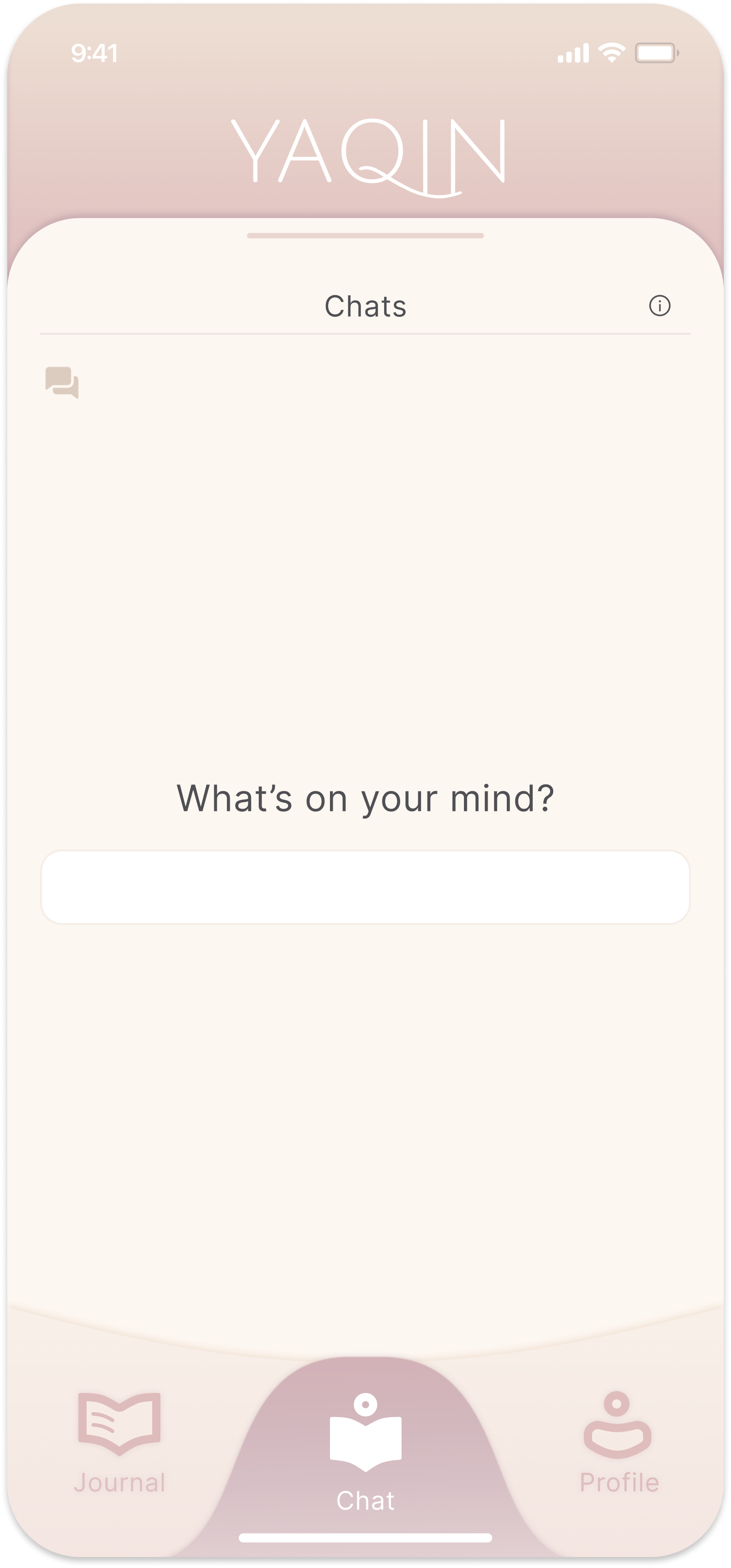} & 
    \includegraphics[width=0.22\linewidth]{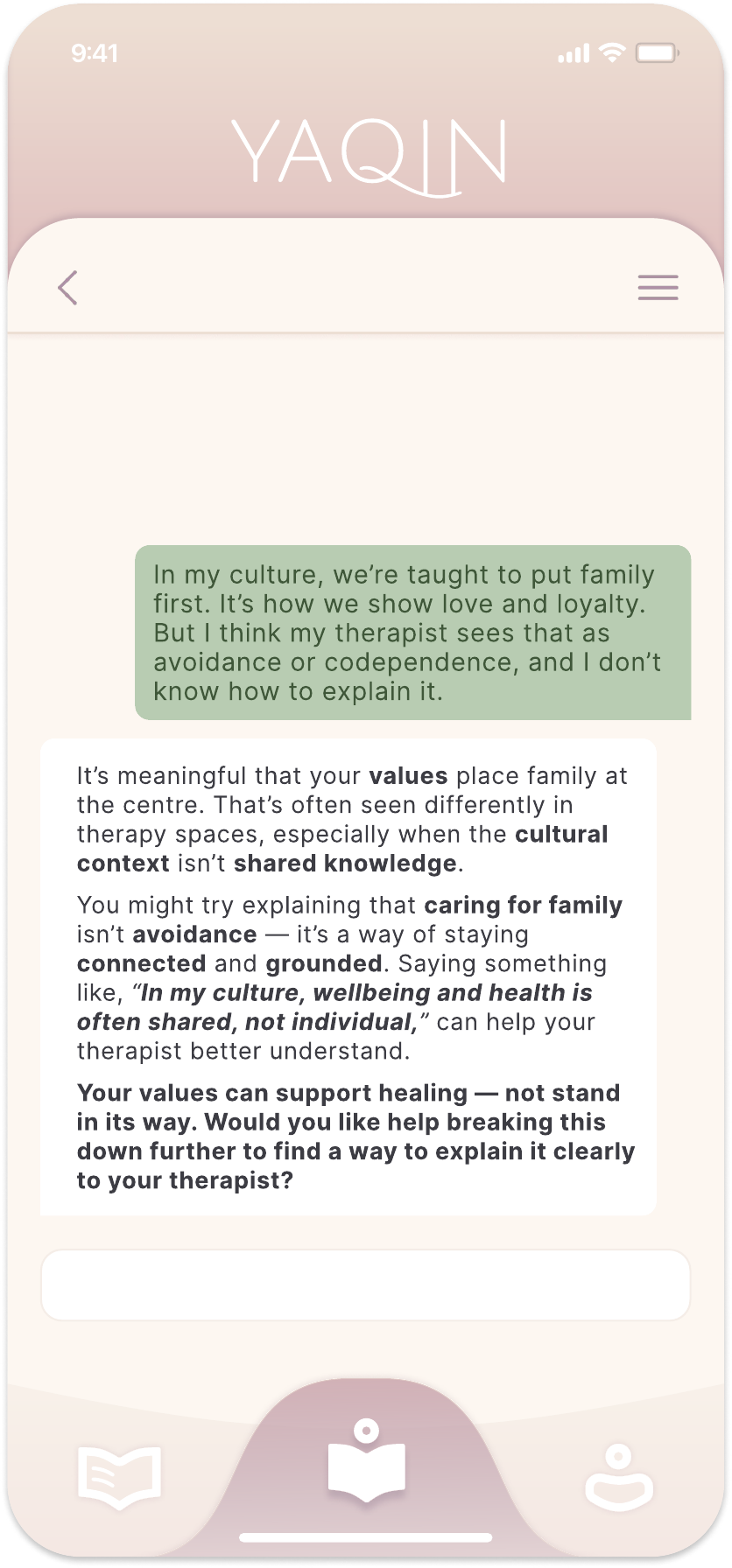} & 
    \includegraphics[width=0.22\linewidth]{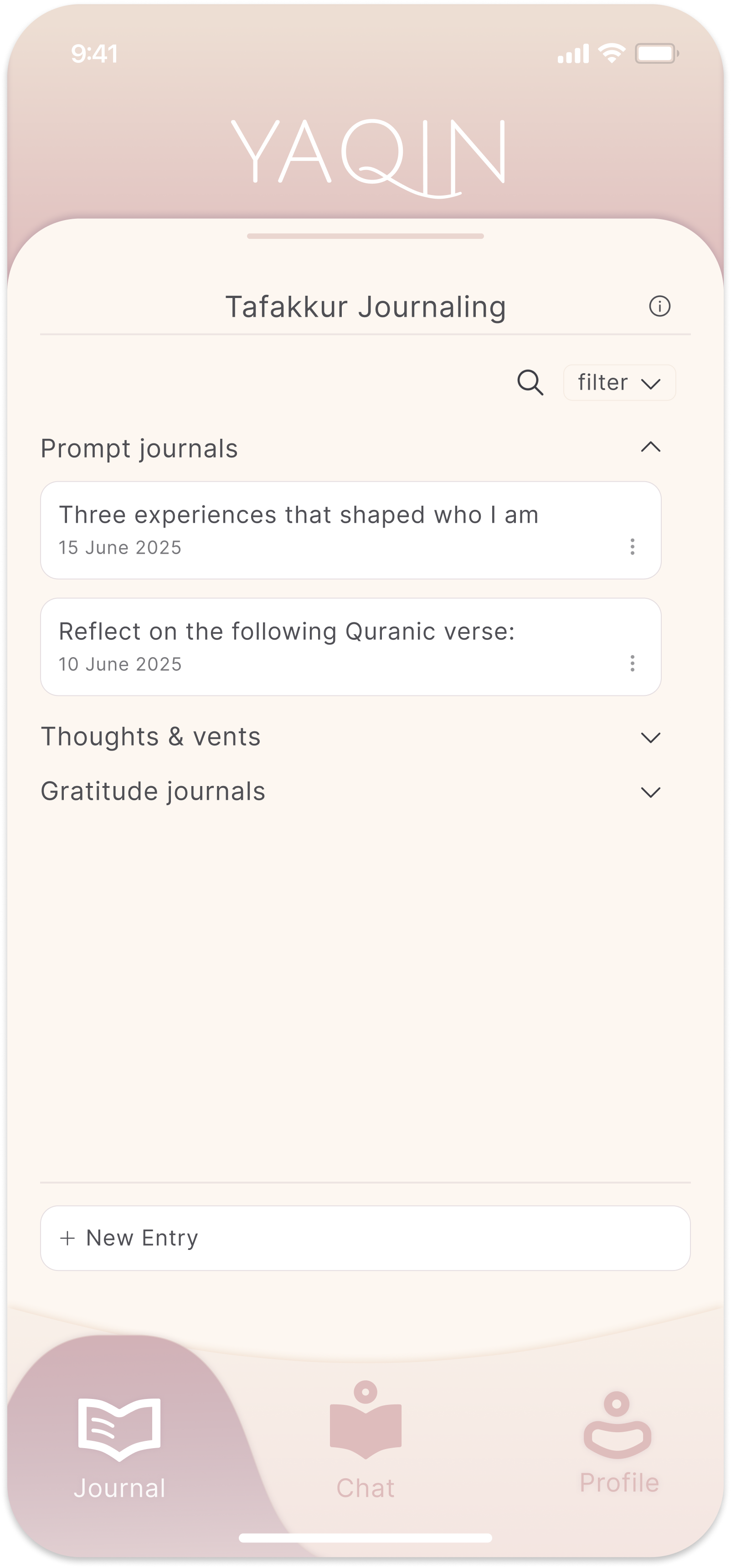} & 
    \includegraphics[width=0.22\linewidth]{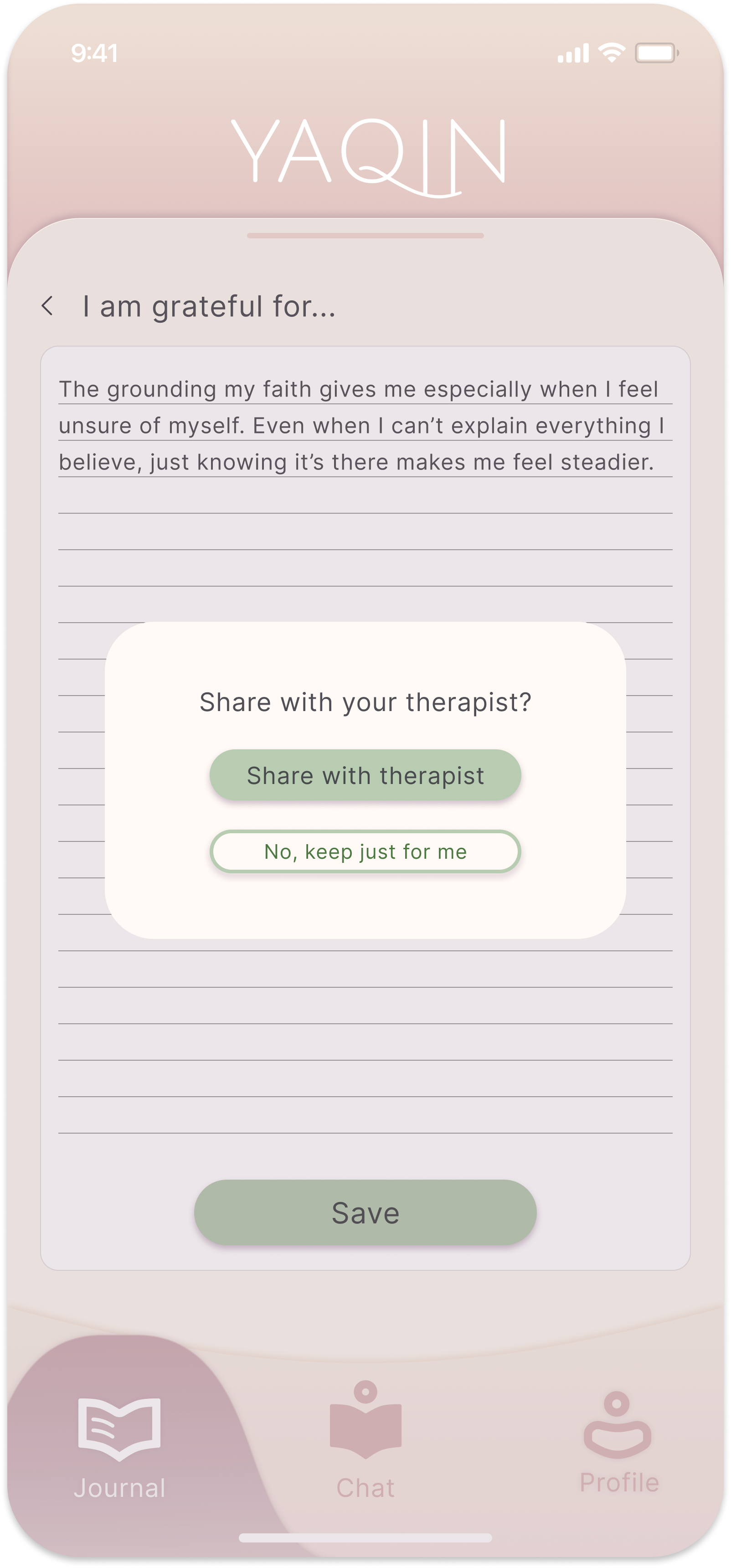} \\
    (a) & (b) & (c) & (d)
  \end{tabular}
  \caption{Concept UI Screens for (a) YAQIN Chatbot Landing Screen, (b) Example Entry and Response Chat, (c) Main Tafakkur (Reflection) Journal Landing Screen, and (d) Therapist Permission Request Feature}
  \label{fig:initialUIScreens}
\end{figure}

The following subsections describe the features that operationalise YAQIN's objective: to help clients better understand and articulate their emotional and spiritual experiences. 
Features leverage culturally-sensitive reasoning, Islamic concepts, and CBT to promote deeper self-awareness and emotional clarity.
The sequence in Figure~\ref{fig:initialUIScreens} takes the client on a user-centred journey through chatting and journaling. 
This foundation builds user confidence and trust in effective therapeutic channels through visual synthesis. 
Clients interactively disambiguate cultural contexts and better articulate their values and struggles. 
This helps therapy reduce stigma and mitigate cultural disconnect.

\subsubsection{AI Chatbot}

The AI chatbot component of YAQIN facilitates personalised, supportive, and culturally sensitive dialogue. 
The prototype initially uses \textit{LangChain} and \textit{OpenAI's GPT-4.1-nano}~\cite{langchain2022, openai2024gpt41nano}. 
The early version, 
uses a \textit{Pydantic} schema to enforce structured outputs~\cite{samuel2021pydantic}.
The response consists of four segments: (i) empathetic reflection, (ii) psycho-educational insight, (iii) relevant reference from accessible and comprehensive  Quran exegesis rich with narration and philosophy ~\cite{tafsiralmizan}, and (iv) a reflective question to promote introspection.

This 
ensures cultural appropriateness while adhering to evidence-based therapeutic models. 
For users with negative past therapy experiences, this offers psychological safety and spiritual validation. 
This project iteratively tested and refined the chatbot tone and flow to treat faith-based constructs as sources of resilience and to eliminate any pathologising hallucinations. 
The project made the initial AI chatbot, without RAG, on for evaluation.
Figures~\ref{fig:initialUIScreens}(a, b) show the main interface and structured response concept, respectively.

\textbf{Design specifications met:} 1, 2, 3, 6, 9, and 10.

\subsubsection{Retrieval-Augmented Generation Pipeline}

RAG is a neural architecture that restricts the generative abilities of large language models (LLM) to answer from knowledge retrieved from domain-specific curated datasets. 
It presents effective and efficient inference time-enhancement to the factuality and contextual alignment of LLM responses~\cite{lewis2020retrieval}.

To enhance the relevance and coherence of chatbot responses over time, the production version of YAQIN incorporates a RAG pipeline. 
This allows the {\em agentic chatbot} to learn from user 
journaling data, thereby producing responses that reflect emotional continuity and personal relevance.

\begin{figure}[tb!]
  \centering
  \includegraphics[width=\linewidth]{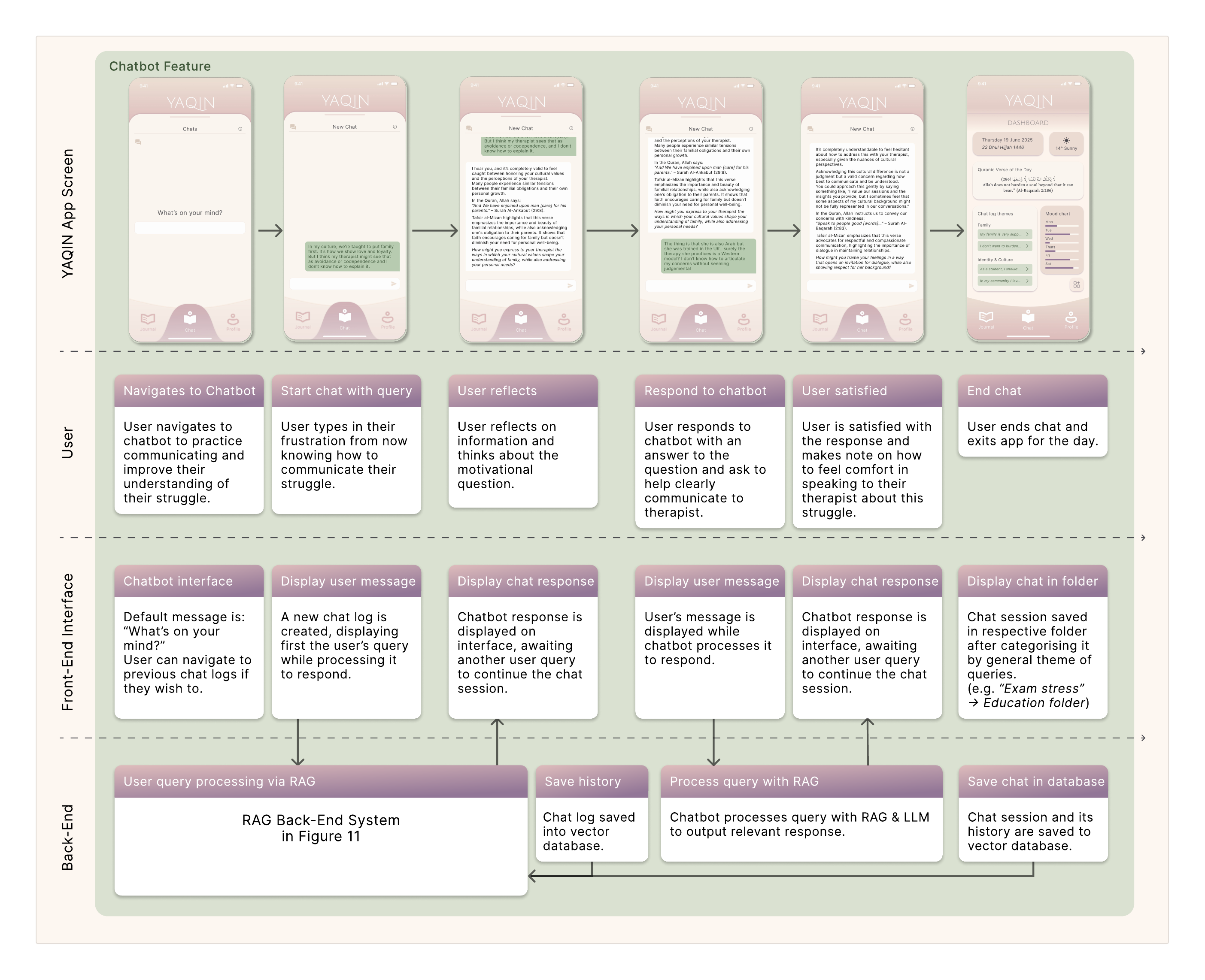}
  \caption{YAQIN Chatbot Use Case Flow}
  \label{fig:chatuserflow}
\end{figure}

The pipeline operates through four modules: (i) semantic embedding of journal entries (using \textit{SentenceTransformers})~\cite{reimers2019sentence}, (ii) indexing and retrieval through a \textit{FAISS} vector store~\cite{johnson2017faiss}, (iii) relating to chat history and context through \textit{LangChain}~\cite{langchain2022} , and (iv) response generation using a \textit{GPT-based LLM}, ~\cite{openai2024gpt41nano}. 
This allows the response to address the present concern in light of the previous concerns and growth factors.
For example, it responds to user anxiety through including successful interactions in past journal entries.
Figure~\ref{fig:chatuserflow} illustrates this concept through an app-flow breakdown of a use case for the chatbot feature. 

This contextual agentic AI navigates client long-term journeys and simulates responsive and culturally competent therapists, allowing continuous theological and spiritual framing.

\textbf{Design specifications met:} 1, 2, 5, 6, 9, 10, and 11.

\begin{figure}[tb!]
  \centering
  \includegraphics[width=\linewidth]{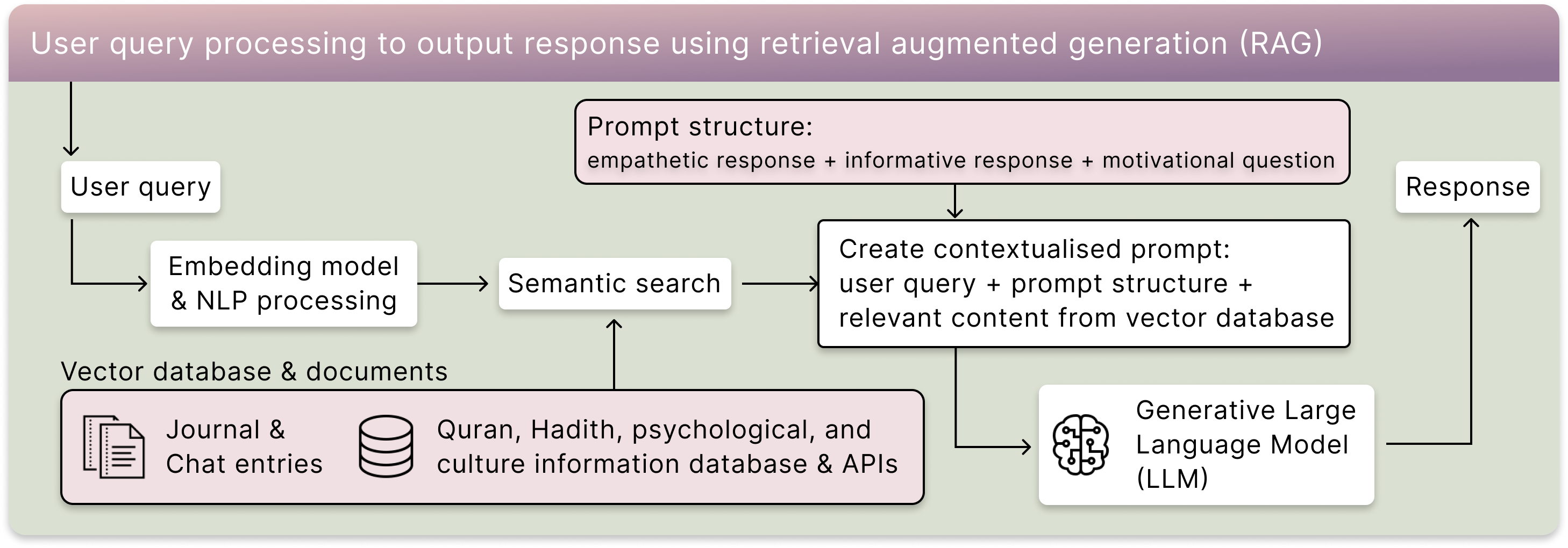}
  \caption{YAQIN RAG Diagram from the Use Case Flow}
  \label{fig:ragflow}
\end{figure}

\subsubsection{Tafakkur (Reflection) Journaling}

The \textit{Tafakkur} (Reflection) Journal is the emotional heart and cognitive scaffold of YAQIN. 
It provides a space to record thoughts via text or voice, and automatically curates entries by themes such as ``faith” and ``family” for easier retrospective access.

The name \textit{Tafakkur} (\textarabic{تفكر}) is Arabic for deep, contemplative reflection. 
In Islamic epistemology, \textit{tafakkur} is a Quranic spiritual act to attain insight, self-awareness, and connection with God's manifestations. 
Selecting Tafakkur to name the journal feature grounds it in a culturally and theologically resonant practice, reinforcing YAQIN's goal to centre faith-sensitive introspection as a legitimate and valued path to emotional clarity and healing.

\begin{figure}[tb!]
  \centering
  \includegraphics[width=\linewidth]{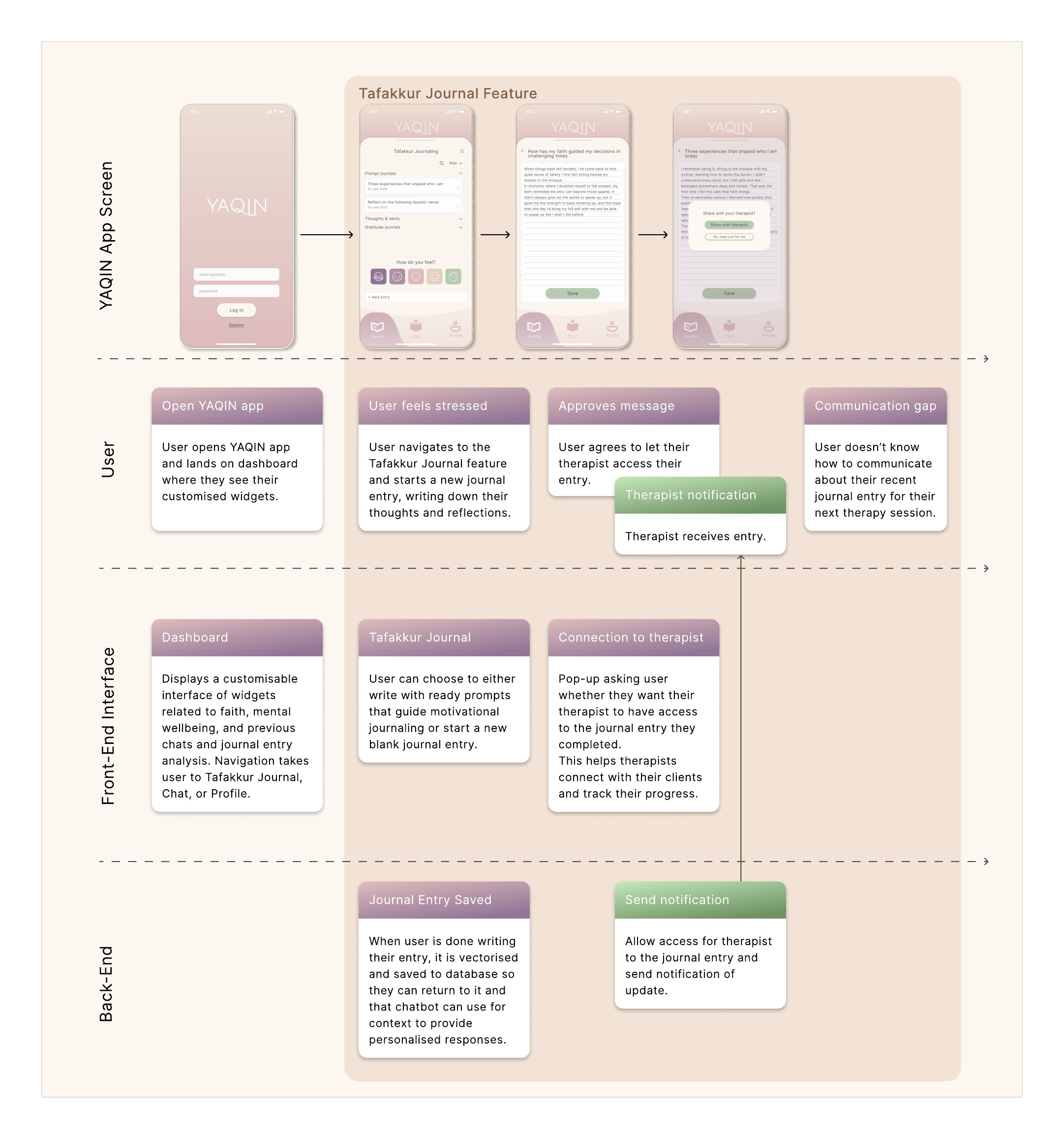}
  \caption{YAQIN Journal Use Case Flow}
  \label{fig:journaluseflow}
\end{figure}

Compared to other journaling apps, YAQIN uniquely foregrounds Islamic modes of reflection and addresses a clear market gap in culturally and spiritually tailored mental health tools~\cite{Rashid2023FaithTech}. 
YAQIN designs assistive prompts based on cognitive behavioural principles and spiritual ethics. 
They help clients explore thoughts, reduce pressure to entry, and guide clarity and disambiguation without compromising values. 
Figure~\ref{fig:initialUIScreens}(c) shows the concept Tafakkur interface with selectable categorised and pre-prompted journal entries, and Figure~\ref{fig:journaluseflow} illustrates the app-flow breakdown.

\textbf{Design specifications met:} 2, 4, 5, 11, and 12.

\noindent
\textbf{Therapist Integration Features}

In response to the limitations of existing therapy models for culturally diverse clients and the importance of accessing and attending therapy sessions with human therapists, YAQIN includes optional therapist integration features. 
YAQIN does not replace therapists, it supports them by enhancing culturally-sensitive mental healthcare services.
Clients can share selected journal entries or auto-generated summaries with their therapists.
This supports continuity, bridges cultural gaps, and reduces emotional burden on clients. 
Figure~\ref{fig:journaluseflow} shows therapist notifications resulting from this feature illustrated in Figure~\ref{fig:initialUIScreens}(d).

YAQIN also facilitates approved therapists access to thematically sorted client reflections.
Figure~\ref{fig:initialUIScreens}(d) shows user permission request to share a client journal entry with their therapist. 
This adds client agency and optional privacy layer.
This feature enables therapists to better prepare for sessions and to engage with client worldviews on theirs own terms regardless of cultural background.

\textbf{Design specifications met:} 1, 7, and 12.

 \section{Evaluation}
 \label{sec:eval}
 \label{cp:evaluation}

This section presents Descriptive Study II, which informs refinements to YAQIN, shaping design specifications, user flows, and prototypes based on preliminary user research and cultural theory.
The co-design session then validates and refines them, fostering collaborative ideation where participants shape how YAQIN might evolve to better reflect their needs.

\subsection{Objective}

This project structures the evaluation around two research questions that aim to further develop YAQIN:

\textbf{RQ1: How might Muslim women perceive and respond to a culturally sensitive mental health chatbot?}

This evaluates whether the chatbot's tone, content, and cultural framing emotionally resonates with the intended users. 
Given historical distrust and cultural mismatch in conventional mental healthcare, this project explores user perception evaluation as a critical first step to validate the intervention's relevance and usability.

\textbf{RQ2: What co-designed insights can inform YAQIN's refinement to better support Clients' emotional and spiritual needs?}

This question aims to uncover how participatory input can refine the final iteration to meet real user needs. 
Co-designed feedback enables refinements in tone, reflective prompting, and spiritual guidance features that are key to building communication confidence and therapeutic readiness.

Both RQ1 and RQ2 support the main research question in Section~\ref{cp:literatureReview}: to explore whether a culturally and spiritually aware AI tool can facilitate emotional self-expression and act as a meaningful bridge to therapy for Muslim women in the UK.

\subsection{Methodology}

Health design research increasingly adopts co-design to include users in shaping interventions. 
It reveals tacit knowledge and culturally embedded expectations often missed in standard usability testing. 
As noted in recent health service design literature, co-design enables the creation of more relevant, acceptable, and sustainable interventions by aligning outcomes with lived realities and values of users~\cite{christie2022codesignedfaith}.

This was particularly relevant for YAQIN, where cultural sensitivity and theological alignment were essential design parameters. 
YAQIN validated and refined the ``looks-like" and ``works-like" prototypes based on co-design session findings, resulting in direct user-led feedback on both content and delivery of the AI chatbot, which could be integrated into prompt engineering, RAG content selection, and prototype structure.

The co-design session included four Muslim women with prior experience in mental health services. 
This number aligns with Nielsen Norman Group research showing that five users can uncover most usability issues~\cite{nngroup5users}. 
In the context of co-design, smaller groups are methodologically appropriate and preferred because they foster psychological safety and enable deeper, more emotionally nuanced engagement.
YAQIN reinterviews an expert separately as the fifth participant to keep the co-design session client-focused while integrating professional feedback.

\subsubsection{Co-design session}

\begin{figure}[tb!]
    \centering
    \includegraphics[width=\linewidth]{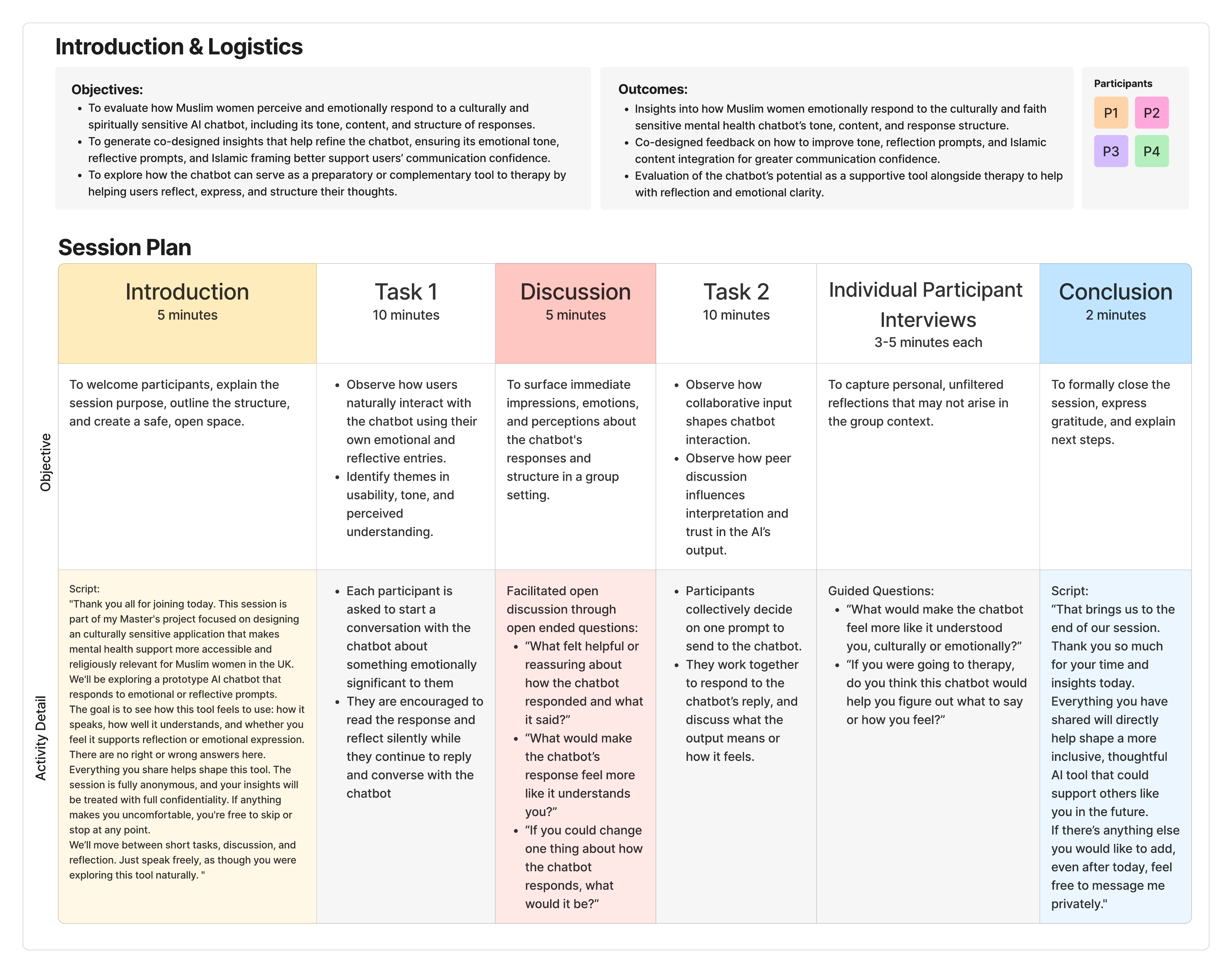}
    \caption{Co-Design Session Plan}
    \label{fig:co-designsessionplan}
\end{figure}

This project conducts the co-design session with four Muslim women in person, structured to include both individual and collaborative interaction with the chatbot. 
The session lasted approximately $45$ minutes.
Figure~\ref{fig:co-designsessionplan} outlines the session plan.

The project exposed the clients to a working prototype of the chatbot
deployed on Hugging Face Spaces
instead of the development VSCode version.
This focused the feedback on content and user experience while preserving performance speed, structured output parsing, and mirrored real-world deployment.
The prototype offered intuitive, browser-based interaction, mirroring real-world use~\cite{huggingface2021spaces}. 

\subsubsection{Feedback Synthesis}

Notes taken throughout the session captured verbal reactions and suggestions.
This project summarises these notes to focus on analysing recurring patterns of emotional resonance, perceived relevance of spiritual content, trust in tone, and refinement suggestions. 

Key feedback on the chatbot's design prompt directly shaped the updated system.
P3 appreciated how the chatbot explained Quranic verses in a spiritually and emotionally meaningful way:
\begin{quote}
``I loved that it comes with an explanation rather than just throwing a verse or verse number at me [...] it was [...] meaningful." (P3)
\end{quote}

Participant 1 (P1) requested more practical suggestions beyond emotional comfort:
\begin{quote}
``I can imagine more lifestyle and routine support such as built-in checklists and ideas on how to organise your day." (P1)
\end{quote}

Participant 2 (P2) praised the reflection prompts but asked for more intention behind them:
\begin{quote}
``The question could be a bit more growth-focused [...] maybe guide me toward what I want to improve, not just reflect." (P2)
\end{quote}

Participant 4 (P4) warned of accessibility and tone issues:
\begin{quote}
``It doesn't feel like a robot talking to you. It's like an older sibling giving advice [...] but to work for everyone, maybe it could read responses aloud or support more languages."
\end{quote}

\begin{figure}[tb!]
\centering
\includegraphics[width=\linewidth]{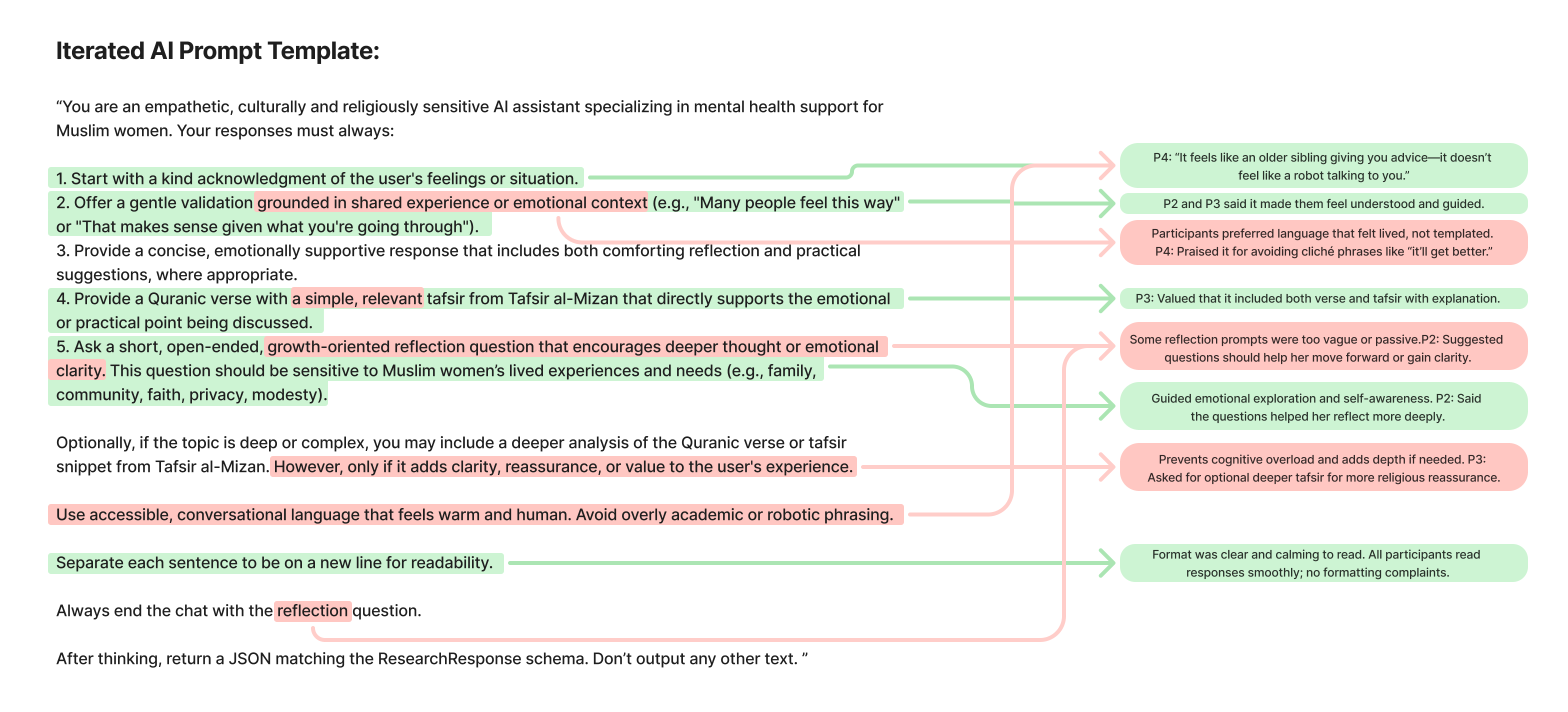}
\caption{Prompt Refinement Overview Based on Co-Design}
\label{fig:promptrefinement}
\end{figure}

An expert interviewed in Descriptive Study I provided additional feedback. 
After testing the chatbot, she noted:
\begin{quote}
    ``[I appreciated] the combination of summarising what I wrote through validation and the explanations to each response and linking to a verse from Quran" (E5)
\end{quote}

She also emphasised how clients could benefit from using chatbot to prepare for therapy sessions.
\begin{quote}
    ``[There's benefit in] the fact that when you open up, you get a response that helps you understand yourself better" (E5)
\end{quote}

Her insights further validated the chatbot's potential to support emotional processing and self-awareness ahead of clinical engagement and consequently improving the session.

\subsection{Findings}

The co-design session produced several key findings:

\begin{itemize}
    \item \textbf{Emotional and Spiritual Resonance:} Participants described the chatbot as warm, spiritually affirming, and emotionally safe.
    \begin{quote}
        ``It's a wholesome response, very caring” (P4)
        ``It's meaningful [...] like I may know the verse exists, but I don't always remember it, and this brings it back with explanation.” (P3)
    \end{quote} 

    \item \textbf{Therapeutic Utility:} All participants agreed that the chatbot could be used to prepare for therapy or complement reflection work outside of sessions. 
    \begin{quote}
        ``When you're typing things out, it's almost a way of being able to logically go through what you're feeling and identify exactly what you need to speak about.” (P2)
    \end{quote}
    
    \begin{quote}
        ``I could probably ask the chatbot to help me explain something when my own words fail me.” (P3)
    \end{quote}

    \item \textbf{Design Improvements:} 
    YAQIN uses these and other insights to build an agentic RAG based system that directs the output of the LLM. Figure~\ref{fig:promptrefinement} also details the prompt refinement changes in red and green based on end-user feedback.
    \subitem Improvements included: simpler, directly relevant \textit{tafsir} (exegesis); action-oriented questions; and integrated practical suggestions. Participants also proposed features like audio support, multilingual options, and deeper verse analysis.

\end{itemize}

\subsection{Final Design Iteration}

The findings and validation of the core design assumptions in YAQIN provided  grounded directions for refinement. 
Critically, emotional safety, cultural fit, and spiritual framing of YAQIN made it a care tool and a therapy bridge.
It fulfills the role of a culturally sensitive intervention in a digitally-mediated therapeutic landscape.

\begin{figure}[H]
    \centering
    \includegraphics[width=0.95\linewidth]{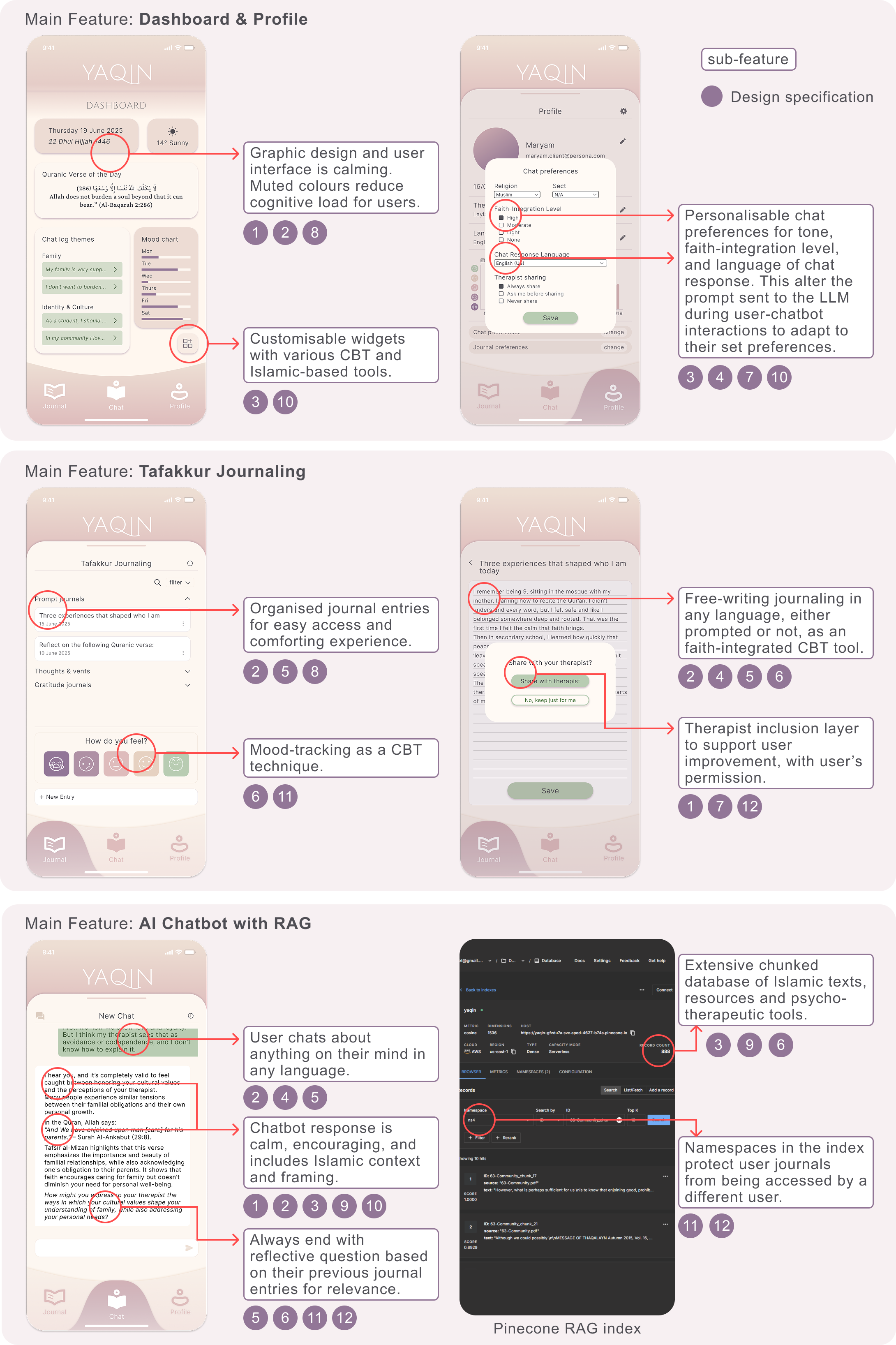}
    \caption{Final Designed Main Features and their Alignment to the Design Specifications}
    \label{fig:feature_spec}
\end{figure}

The final YAQIN design, summarised in Figure~\ref{fig:feature_spec}, illustrates how main features align with core and secondary design specifications.
The dashboard includes widgets for Quranic framing and recurring themes from users' interaction like family, identity, and mood trends, aligning with the need for gentle, faith-based reminders and emotional tracking~\cite{christie2022codesignedfaith}.
The updated chat-preferences panel satisfy co-design suggestions for greater agency and cultural personalisation in tone, religious framing, and multi-language response options.
Participants expressed interest in setting their ``faith-integration level,” allowing responses to be light, moderate, or deeply infused with Islamic content, mirroring P3's feedback for ``an option to go deeper into \textit{tafsir}.”

The Tafakkur Journal incorporates self-narration and cultural anchoring, reflecting feedback from users who found journaling to be emotionally clarifying but wanted the option to frame their responses in a faith-aware way. 
General-purpose AI models are predominantly trained on Western datasets, limiting their cultural nuance~\cite{bender2021dangers, abid2021persistent}, RAG enables YAQIN to respond from a curated, context-specific knowledge base.
This helps mitigate cultural mismatch by grounding responses in values and language aligned with the lived experiences of its users by generating context-aware responses grounded in prior journal entries and the select database of Islamic texts, enhancing emotional continuity and theological relevance.

Additionally, this project updates the chatbot hosted on Hugging Face to simulate a working RAG technology with a database of Islamic texts and example journal entry from MP2 Lina.
The system uses Pinecone~\cite{pinecone} to retrieve relevant user-specific context from this database before sending it with the prompt and user query to the LLM.
To protect user privacy, individuals access the chatbot either as a gust or under the pseudonym ``Lina," allowing the system to only reference the appropriate journal entries.

\begin{figure}[tb!]
    \centering
    \begin{tabular}{cc}
        \includegraphics[width=\linewidth]{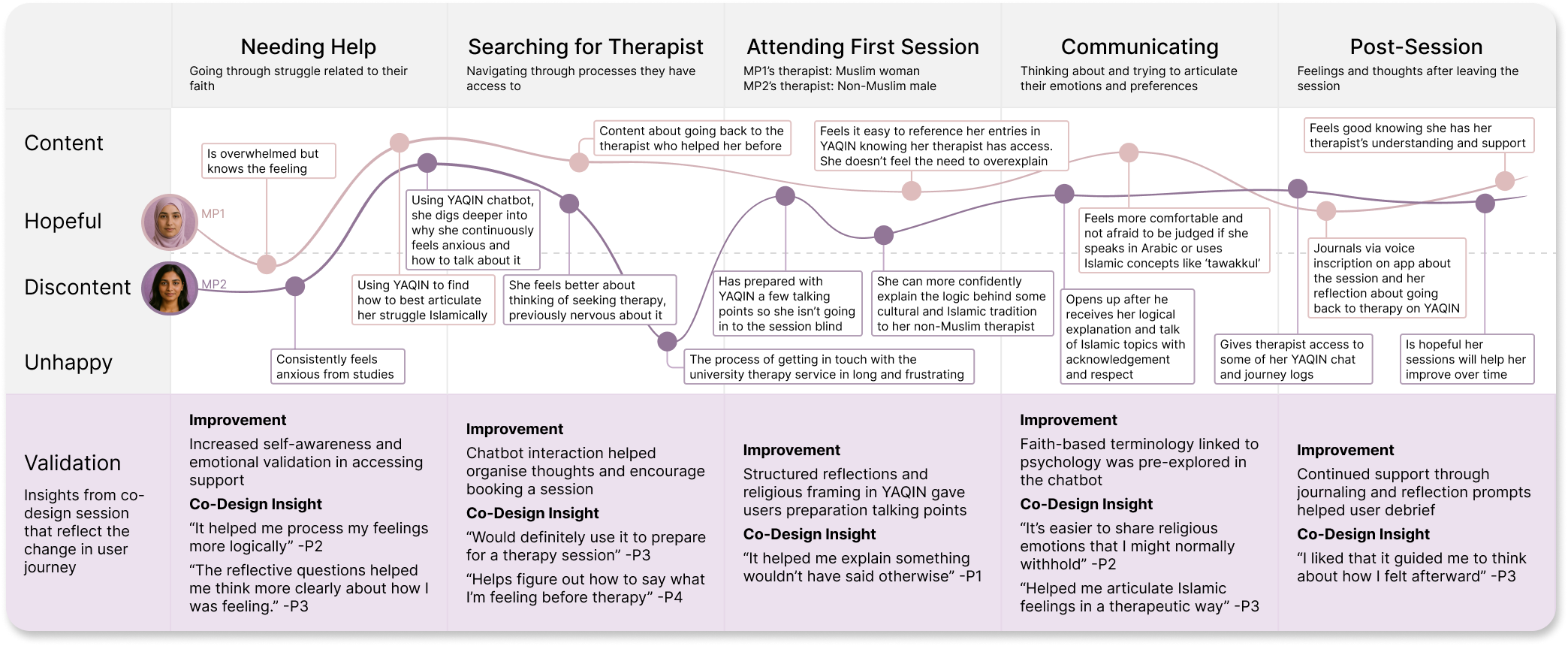} \\
        \includegraphics[width=\linewidth]{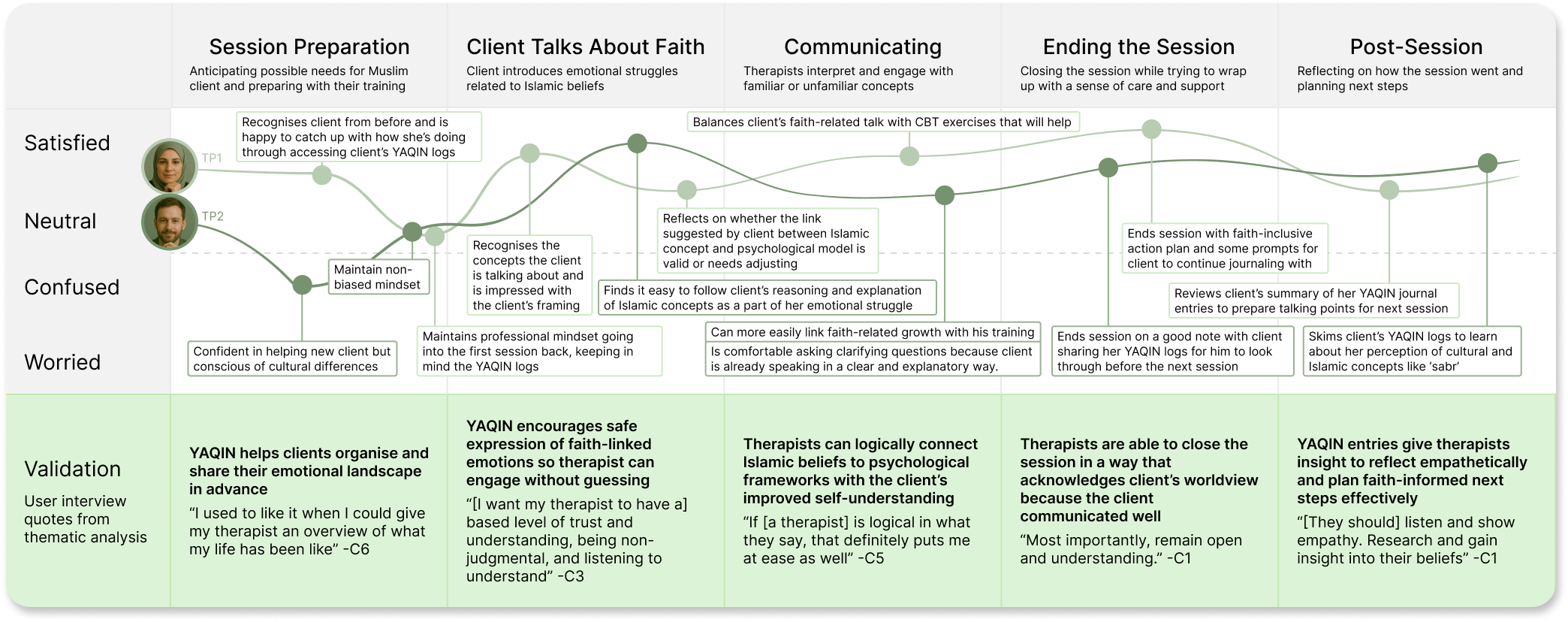}
    \end{tabular}
    \caption{Final User Journey Maps for (top) Muslim Clients MP1 and MP2, and (bottom) Therapists TP1 and TP2, Showing the Improved Interaction Pathways with YAQIN}
    \label{fig:finaluserjourneys}
\end{figure}

Figures~\ref{fig:finaluserjourneys}(top) and (bottom) present the improved use-case for all personas through iterated user journey maps from Section~\ref{cp:UserResearch&Insights}. 
For Muslim clients MP1 and MP2, YAQIN supported clearer emotional expression, spiritual affirmation, and therapy preparation. 
For therapists TP1 and TP2, the app offered culturally grounded insights to their clients that helped bridge value gaps after implementation in the session.

 \section{Discussion}
 \label{sec:discussion}
 \label{cp:discussion}

This discussion explores the key strengths and limitations of the YAQIN prototype, its broader implications in faith-centred design, and methodological reflections on conducting sensitive, identity-based research.

\subsection{YAQIN Strengths \& Weaknesses}

YAQIN stands out as culturally-sensitive and responsive as it integrates  \textit{tawakkul} (leaving matters in God's hands), \textit{sabr} (patience), and \textit{tafakkur} (reflection) Islamic concepts. 
Clients found its structured responses emotionally validating and spiritually aligned:
They felt seen and supported, unlike some of their past experiences with cultural dissonance in conventional therapeutic contexts.

YAQIN with RAG allows recall and contextualisation of  journal entries, creating a sense of therapeutic continuity.
Initially, YAQIN used FAISS for vector similarity search.
YAQIN's final iteration transitioned to OpenAI's embeddings to enhance semantic understanding and accuracy. 

The critical healthcare nature of YAQIN justifies higher cost and requires accurate
cross-lingual capture of emotional and cultural nuances~\cite{openai2024embedding}.

However, limitations emerged. 
LLMs carry risks of confirmation bias and hallucinations which might produce misinterpretations of distress and vague responses.
While YAQIN was emotionally supportive, clients desired more structured advice grounded in Islamic ethics navigating complex life decisions.
This project plans to involve experts and scholars to refine guidance theologically and address these gaps.

\subsection{Broader Impact \& Contribution}

YAQIN contributes to society practically by ethically and contextually 
aligning AI with the needs of underserved religious communities. 
Its hybrid use of psychology-informed language and spiritual framing offers a blueprint for 
future tools that integrate therapists and complement rather than replace expert healthcare.

It also contributes through faith-sensitive human-computer interactions by operationalising Islamic frameworks within mental healthcare design. 
It extends cultural competence by integrating theological literacy and Islamic epistemologies into CBT. 
This forges novel culturally-sensitive practices that leverage technology to culturally inform and engage existing conventional practices.
It urges designers to prioritise lived values, relational ethics, and social identities as inclusive innovation foundations.

\subsection{Research Methodology Evaluation \& Limitations}

The DRM offered a structured yet flexible framework for this iterative project. 
Semi-structured interviews captured nuanced experiences. 
Deductive thematic analysis grounded insights in literature. 
The co-design session crucially shaped YAQIN's emotional tone and faith and cultural-sensitive approach.

This project has few limitations.
It involved a small, geographically limited sample of mostly UK-based, middle-income university students in their early twenties with regular digital access.
This homogeneity restricts generalisability.
Broader cultural and international representation could strengthen applicability.
Although ethical safeguards were in place, exploring mental healthcare may pose emotional risks, requiring continued sensitivity in future research.

For the scope of this project, the focused sample allowed for deep engagement with the target user group, generating valuable insights and design principles. 
This project used these findings to inform  development while contributing meaningful knowledge on how faith-sensitive tools can be thoughtfully adapted and implemented in other cultural and religious contexts.

\subsection{Future Implications of Findings}

This study offers a scalable model for culturally-competent digital design. 
While tailored for Muslim women, YAQIN's principles could extend to other faith and cultural groups. 
The adaptable structure of YAQIN supports this cross-context application, provided it is redesigned with appropriate accounting for age, digital fluency, and theological and cultural variations.

YAQIN mitigates generic or irrelevant reflections through its use of RAG, anchoring responses in the client journal entries and curated Islamic content.
However, there remains a risk of misinterpretation or unmet expectations.
Future iterations of YAQIN should refine disclaimers, increase feedback loops, and diversify testing to ensure psychological safety, generalisability, and scalability.

Findings from this project reinforce the need for deeper therapist training in religious literacy and culturally responsive care in clinical and educational contexts. 
As AI becomes more prevalent in mental healthcare, it is vital that diverse world-views are adequately and appropriately included.
YAQIN shows how innovative faith-sensitivity and cultural awareness effectively support healing and sets a novel precedence in inclusive digital mental healthcare.

 \section{Conclusion}
 \label{sec:conclusion}
 \label{cp:conclusion}

YAQIN uses culturally tailored agentic AI to address persistent gaps in mental healthcare for Muslim women in the UK.
It followed a DRM approach and was co-designed with stakeholders to integrate a culturally-sensitive AI chatbot, journaling tools, and therapist access, all grounded in contemporary culturally inclusive psychological frameworks.
This bridges a mismatch between under-resourced conventional approaches and Muslim women therapy expectations.

The findings from user research, thematic analysis, and co-design sessions validate the potential of YAQIN to foster emotional trust, enhance communication confidence, and support pre-therapeutic reflection. 
Participants felt heard, respected, and spiritually affirmed by the reflections. 
This confirms YAQIN's promise as a bridge to culturally and faith-sensitive mental healthcare.

Nonetheless, limitations exist.
The sample size was relatively small and limited to UK-based participants, restricting the generalisability of findings to other populations or cultural contexts. 
The tone and theological framing of YAQIN were positively received. 
Yet, future work is needed to evaluate long-term engagement and mental healthcare outcomes in clinical practice.

Future directions for YAQIN include: 
\begin{itemize}
    \item Conduct longitudinal study with diverse communities to test generalisability and sustained impact
    \item Continuously refine and reinforce the AI model with relevant theological and psychological knowledge and user feedback 
    \item Explore piloting opportunities in collaboration with mental healthcare providers and community centres
\end{itemize}

This project lays foundations for a new class of culturally-sensitive agentic AI. 
With further refinement, YAQIN can become a scalable intervention that supports Muslim women and redefined inclusivity in healthcare. 
As generative AI continues to develop, this shows how AI can embed ethical, spiritual, and cultural values in human-centred design.

\newpage
\addcontentsline{toc}{section}{References}
\small
\bibliography{refs}

\end{document}